\documentclass[aps,oneclumn,notitlepage,superscriptaddress,nofootinbib]{revtex4-1}
\usepackage{graphicx}
\usepackage{amssymb,amsmath,bm}
\usepackage{mathtools}
\usepackage[english]{babel}
\usepackage{epsfig}
\usepackage{epstopdf}
\usepackage{subfig}

\usepackage{float}
\usepackage{natbib}
\usepackage{siunitx}
\usepackage{hyperref}
\usepackage{placeins}
\usepackage{color}

\newcommand{\COM}[1]{\textcolor{black}{#1}}
\newcommand{\CR}[1]{\textcolor{black}{#1}}

\begin{document}

\title{Sharp transitions in rotating turbulent convection: Lagrangian acceleration statistics reveal a second critical Rossby number}

\author{Kim M. J. \surname{Alards}}
\author{Rudie P. J. \surname{Kunnen}}
\affiliation{Fluid Dynamics Laboratory and J.M. Burgers Center for Fluid Dynamics, Department of Applied Physics, Eindhoven University of Technology, P.O. Box 513, 5600 MB Eindhoven, The Netherlands}
\author{Richard J. A. M. \surname{Stevens}}
\affiliation{Physics of Fluids Group, Max Planck Center for Complex Fluid Dynamics, J.M. Burgers Center for Fluid Dynamics and MESA+ Research Institute, 
Department of Science and Technology, University of Twente, P.O. Box 217, 7500 AE Enschede, The Netherlands}
\author{Detlef \surname{Lohse}}
\affiliation{Physics of Fluids Group, Max Planck Center for Complex Fluid Dynamics, J.M. Burgers Center for Fluid Dynamics and MESA+ Research Institute, 
Department of Science and Technology, University of Twente, P.O. Box 217, 7500 AE Enschede, The Netherlands}
\affiliation{Max Planck Institute for Dynamics and Self-Organization, Am Fassberg 17, 37077 G{\"{o}}ttingen, Germany}
\author{Federico \surname{Toschi}}
\affiliation{Fluid Dynamics Laboratory and J.M. Burgers Center for Fluid Dynamics, Department of Applied Physics, Eindhoven University of Technology, P.O. Box 513, 5600 MB Eindhoven, The Netherlands}
\affiliation{Centre of Analysis, Scientific Computing, and Applications W\&I, Department of Mathematics and Computer Science, Eindhoven University of Technology, P.O. Box 513, 5600 MB Eindhoven, The Netherlands}
\affiliation{Istituto per le Applicazioni del Calcolo, Consiglio Nazionale delle Ricerche, Via dei Taurini 19, 00185 Rome, Italy}
\author{Herman J. H. \surname{Clercx}}
\email{h.j.h.clercx@tue.nl}
\affiliation{Fluid Dynamics Laboratory and J.M. Burgers Center for Fluid Dynamics, Department of Applied Physics, Eindhoven University of Technology, P.O. Box 513, 5600 MB Eindhoven, The Netherlands}
\begin{abstract}
\noindent In Rayleigh--B{\'{e}}nard convection (RBC) for fluids with Prandtl number $Pr\gtrsim 1$, 
{\CR{rotation beyond a critical (small) rotation rate}} is known to cause a sudden enhancement of heat transfer 
which can be explained by a change in the character of the boundary layer (BL) dynamics near the top and bottom plates of the convection cell. 
Namely, with increasing rotation rate, the BL signature suddenly changes from Prandtl--Blasius type to 
Ekman type. The transition from a constant heat transfer to an 
almost linearly increasing heat transfer with increasing rotation rate is known to be 
sharp and the critical Rossby number $Ro_{c}$ occurs typically in the range $2.3\lesssim Ro_{c}\lesssim 2.9$ (for Rayleigh number $Ra=1.3\times 10^9$, $Pr=6.7$, and a convection cell with aspect 
ratio $\Gamma=\frac{D}{H}=1$, with $D$ the diameter and $H$ the height of the cell). The explanation 
of the sharp transition in the heat transfer points to the change in the dominant flow structure. At $1/Ro\lesssim 1/Ro_c$ (slow rotation), 
the well-known large-scale circulation (LSC) is found: a single domain-filling convection roll made up of 
many individual thermal plumes. At $1/Ro\gtrsim 1/Ro_c$ (rapid rotation), the LSC vanishes and is replaced with 
a collection of swirling plumes that align with the rotation axis. In this paper, by numerically studying Lagrangian acceleration statistics, related to the 
small-scale properties of the flow structures, we reveal that this transition between these different 
dominant flow structures happens at a second critical Rossby number, $Ro_{c_2}\approx 2.25$ (different from 
$Ro_{c_1}\approx 2.7$ for the sharp transition in the Nusselt number $Nu$; both values for the parameter 
settings of our present numerical study). 
{\CR{When statistical data of Lagrangian tracers near the top plate are collected, it is found that the root-mean-square (rms) values}} and the kurtosis of 
the horizontal acceleration {\CR{of these tracers}} show a sudden increase at $Ro_{c_2}$. To 
better understand the nature of this transition we compute joint statistics of the Lagrangian 
velocity and acceleration of fluid particles and vertical vorticity near the top plate. It is 
found that for $Ro \gtrsim 2.25$ there is hardly any correlation between the vertical vorticity and extreme 
acceleration events of fluid particles. For $Ro \lesssim 2.25$, on the other hand, vortical regions are much more 
prominent and extreme horizontal acceleration events are now correlated to large values of positive (cyclonic) 
vorticity. This suggests that the 
observed sudden transition in the acceleration statistics is related to thermal plumes with cyclonic 
vorticity developing in the Ekman BL and subsequently becoming mature and entering the bulk of the flow for $Ro \lesssim 2.25$. 
\end{abstract}
\maketitle
\section{Introduction}

Turbulent flows in nature are often driven by temperature gradients, for example, oceanic currents 
\cite{Gascard2002,Marshall1999} or large-scale flows in the atmosphere \cite{Hadley1753,Emanuel1994}. A well 
known set-up for studying thermally driven turbulence is Rayleigh--B\'{e}nard convection (RBC)~\cite{Ahlers2009,Lohse2010,Chilla2012}, where a 
confined fluid layer is heated from below and cooled from above. In general, turbulent flows are 
characterized by random fluctuations, intermittency and a loss of temporal and spatial coherence \cite{Pope2000}. 
Sudden transitions between turbulent states are therefore unexpected in strongly turbulent flows. 
In RBC, however, sudden transitions between 
different turbulent states do occur when the set-up is rotated about its vertical axis (see, 
e.g,~\cite{Liu1997,Kunnen2008,Stevens2009,Zhong2009,Kunnen2010,Weiss2010,Zhong2010,Stevens2011,Wei2015,Rajaei2016,Chong2017}). 
These turbulent states are typically characterized by different large-scale coherent flow structures and 
different heat transfer properties. Not only in rotating RBC, but also in Taylor--Couette and 
Von K\'{a}rm\'{a}n flows transitions between different turbulent states are observed 
\cite{Lewis1999,Grossmann2016,Ravelet2004,Ravelet2008}. In a Taylor--Couette flow, i.e. the flow between two 
concentric co- or counter-rotating cylinders, a transition to a different scaling regime sets in when the 
boundary layers (BLs) become turbulent. This is known as the ultimate regime \cite{Grossmann2016}. It was recently shown that even beyond this transition at high 
Reynolds numbers multiple states of turbulence are possible \cite{Grossmann2016,Huisman2014}. In a 
Von K\'{a}rm\'{a}n flow generated between two counter-rotating disks, bifurcations between turbulent states 
occur, which are characterized by different coherent flow structures \cite{Ravelet2004,Ravelet2008}. Like in 
the Taylor--Couette flow, these states can coexist at high Reynolds numbers \cite{Ravelet2004}. 

Here, we focus on the transition in rotating RBC (in convection cells with aspect ratio 
$\Gamma=\frac{D}{H}=1$, with $D$ the diameter and $H$ the height of the cell) from a rotation-unaffected regime, where the heat transfer is 
constant, to a rotation-affected regime where the heat transfer is enhanced \cite{Liu1997,Kunnen2006,Kunnen2008,Rossby1969,Stevens2010a,Stevens2009,Vorobieff1998,Zhong2009}. 
At this transition, the boundary layers change from the passive Prandtl--Blasius type to the active 
Ekman type \cite{Kunnen2011,Rajaei2016,Stevens2010b}. The flow structures change drastically, from the 
domain-filling large scale circulation (LSC) in absence of rotation (or very mild rotation rates) to the 
emergence of a collection of vertically-aligned vortical plumes \cite{Julien1996,Kunnen2008,Rajaei2016a} 
at higher rotation rates. Exactly this transition in the BL dynamics is expected to be 
responsible for the increase in the heat transfer efficiency: in rotating flows strong vortical plumes 
emerge from the Ekman BL, transporting warm (cold) fluid from the BL at the bottom (top) plate into the bulk 
flow and enhancing the heat transfer \cite{Rossby1969,Julien1996a,Kunnen2006,Stellmach2014,Stevens2009}. This mechanism 
is referred to as Ekman pumping. The traditional view \cite{Stevens2013} is that the dominant transition in the flow structure 
(from LSC to vertically-aligned vortices) happens at $Ro_c$. Here we show that this transition takes place 
at a different Rossby number, here denoted by $Ro_{c_2}$. 

It is thus expected that the dynamics of thermal plumes emerging from the BLs majorly determines the heat transfer 
efficiency. In the Prandtl--Blasius type BL, sheet-like plumes develop, while in the Ekman type BL vortical plumes 
emerge, characterized by spiraling fluid motion inside the vortex tubes. The plume dynamics in both the 
rotation-unaffected and the rotation-affected regime has been investigated before. Few of these studies focused 
on the vorticity signature of these plumes, and in particular the presence of vertical vorticity. For example, in 
\cite{Shishkina2008} geometrical characteristics of the sheet-like 
plumes emerging from the Prandtl--Blasius BL are studied and it is shown that these plumes are typically characterized 
by large positive and large negative values of the vertical vorticity. In the rotation-affected regime, on the contrary, 
vortical plumes are characterized by cyclonic (i.e., positive) vorticity only and spiraling fluid motions near the 
horizontal plates \cite{Julien1996,Vorobieff1998,Kunnen2006}. Vertical vorticity might thus be used to 
distinguish the plume character outside the BL for the two regimes.

{\CR{For completeness it is worthwhile to mention that another transition exists in rapidly-rotating 
RBC, which should not be confused with the transition being considered in this investigation: the transition from the rotation-affected 
to the rotation-dominated (or geostrophic) regime \cite{King2012,Julien2012,Horn2014,Ecke2014,Rajaei2018}. 
Knowledge about this regime is still limited. Three main mechanisms have been proposed to explain 
this transition. King {\it{et al.}}~\cite{King2012} hypothesized that the transition from the 
rotation-affected to the geostrophic regime depends on the relative thickness of the viscous and 
thermal boundary layers, $\delta_{\nu}$ and $\delta_T$, respectively. With water as working 
fluid (with $Pr>1$) they showed that $\delta_{\nu}\gtrsim \delta_T$ for the rotation-affected 
regime, and $\delta_{\nu}\lesssim \delta_T$ for the geostrophic regime. The transition between both regimes then 
occurs when $\delta_{\nu}\approx \delta_T$. Julien {\it{et al.}}~\cite{Julien2012} suggested 
that the transition occurs when the vortical plumes span throughout the entire domain (with the 
bulk becoming fully rotation-dominated) resulting in suppression of vertical motion and thus reduction of heat transfer. Recently, Rajaei and coworkers~\cite{Rajaei2018} 
provided evidence that the transition from rotation-affected to the geostropic regime occurs when fluid motion in 
the vertically-aligned vortices becomes decorrelated from the bulk fluid motion between these vortical plumes.}}

\begin{table}
\caption{\textit{Several values for the critical Rossby number $Ro_c$ as found in numerical simulations~\cite{Kunnen2008,Stevens2009,Stevens2011} and 
experiments~\cite{Stevens2009,Zhong2009,Zhong2010,Weiss2010,Stevens2011,Wei2015}. Note 
that the estimated value for $Ro_{c}$ reported by Wei~{\it{et al.}}~\cite{Wei2015} is based on 
the same dataset as the one reported by Zhong and Ahlers for the case with $Pr=4.38$ \cite{Zhong2009,Zhong2010}. For all these cases 
convection cells with $\Gamma=1$ are considered. The critical Rossby 
number is an estimate obtained from each of these studies and these data indicate that $2.3\lesssim Ro_{c}\lesssim 2.9$ 
(for the range of Rayleigh and Prandtl numbers considered in these studies).}}
\centering
\centering
\begin{tabular}{ l c c c l }
\hline\hline\noalign{\smallskip}
\vspace{0.1cm}
${\rm{Reference}}$ &~~~~~&$Ra$ &~~~$Pr$&~~~$Ro_{c}$\\
\hline
Kunnen~${\it{et~al.}}$~\cite{Kunnen2008}&~~~~~&$1.0\times 10^9$&~~~6.4&~~~2.5\\
Stevens~${\it{et~al.}}$~\cite{Stevens2009}&~~~~~&$2.73\times 10^8$&~~~6.26&~~~2.6\\
Zhong~${\it{et~al.}}$~\cite{Zhong2010}&~~~~~&$2.19\times 10^9$&~~~6.26&~~~2.9\\
Weiss~${\it{et~al.}}$~\cite{Weiss2010}&~~~~~&$2.25\times 10^9$&~~~4.38&~~~2.4\\
Stevens~${\it{et~al.}}$~\cite{Stevens2011}&~~~~~&$2.91\times 10^8$&~~~4.38&~~~2.5\\
Zhong~${\it{et~al.}}$~\cite{Zhong2009,Zhong2010}&~~~~~&$2.25\times 10^9$&~~~4.38&~~~2.4\\
Wei~${\it{et~al.}}$~\cite{Wei2015}&~~~~~&$2.3\times 10^9$&~~~4.38&~~~2.3\\
Present work&~~~~~&$1.30\times 10^9$&~~~6.70&~~~2.7\\
 \hline
\end{tabular}
\label{Tab1}
\end{table}

The transition in heat transfer efficiency {\CR{when passing from the rotation-unaffected 
to the rotation-affected regime has been explored in great detail}} in recent years with laboratory experiments 
and numerical simulations. A summary of the main results with respect to the critical Rossby number (the ratio 
of the inertial force to the Coriolis force) for the transition is provided in Table~\ref{Tab1}. It can be 
concluded that $2.3\lesssim Ro_{c}\lesssim 2.9$ for the range of Rayleigh numbers $Ra$, which is a measure 
for the ratio of buoyancy to viscous forces, and the Prandtl numbers $Pr$, the ratio of kinematic 
viscosity to the thermal diffusion coefficient, typical for laboratory 
experiments on turbulent convection with water as working fluid. Although the value of $Ro_{c}$ varies a bit from one experiment to 
another and also differs somewhat 
when compared to simulations (the precise values are also affected by the particular method to determine 
this critical Rossby value, which are relatively rough estimates in many cases), one of the most remarkable 
observations is the sharpness of this transition, see in particular Refs.~\cite{Wei2015,Weiss2010} and the other cited works in 
Table~\ref{Tab1}. Although the transition in the heat transfer due to Ekman pumping is accepted to be sharp, 
it is not yet clear whether and how this transition is reflected in the modified flow structures 
(from LSC to vertically-aligned vortices) in general, 
the plume dynamics in particular, and the emergence and penetration of these structures in the bulk just 
outside the BLs. Moreover, there is no a priori reason to expect a change in the dominant flow structures at exactly the 
same rotation rate where the change in heat transfer efficiency is observed. The heat transfer is, moreover, 
an integral quantity, while flow structures are characterized and quantified by local flow properties, for example, fluid particle 
velocity and acceleration (and its higher order statistics), geometrical properties of tracer trajectories, 
(vertical) vorticity, etc. However, since the heat transfer is expected to be 
related to the typical flow structures and plume dynamics, one might expect that also these local flow quantities 
are affected by the transition. The question is whether this transition in the behavior 
of local flow quantities occurs at the same critical Rossby number $Ro_c$, or does it require a somewhat 
different rotation rate to suddenly change the behavior of such local flow quantities? Alternatively, one could imagine that 
the transition proceeds more gradually outside the BLs as the vortical plumes can gradually penetrate further into the bulk with 
increasing rotation.\\
\indent To study coherent structures in turbulent flows, a Lagrangian approach has shown to be particularly useful 
\cite{Green2007,Schumacher2009}. An interesting quantity is the acceleration of passive tracers, providing 
information on the small-scale flow structures and the temporal fluctuations in these flow structures. High 
acceleration events typically occur in intense vortex filaments \cite{Choi2010,Biferale2004,Moisy2004} and 
are correlated to small-scale intermittency \cite{Schumacher2009,Voth2002}. Rotation is changing the flow 
structures in turbulent flows both at large and small scales \cite{Rajaei2016a} and is therefore expected 
to modify the acceleration statistics. The effect of rotation on the (small-scale) flow structures in 
rotation-affected RBC is most prominent near the plates \cite{Rajaei2016a}. In this region of the flow thermal plumes develop 
from the BLs, resulting in highly accelerating fluid parcels. The nature of the thermal plumes changes with the 
BL transition under rotation \cite{Julien1996a}, from sheet-like in the weakly rotating 
regime to vertically-aligned vortical plumes in the rotation-affected regime, and a signature of this 
transition is certainly expected to be visible in the Lagrangian acceleration statistics in rotating RBC.\\
\indent Acceleration statistics of passive tracers in rotating turbulence have recently been studied experimentally 
\cite{Castello2011,Rajaei2016}. In \cite{Castello2011}, rotation was found to widen the tails of the 
horizontal acceleration PDFs in the bulk of isothermally forced rotating turbulence, while it suppresses the intermittency of the vertical Lagrangian 
acceleration statistics, i.e. in the direction parallel to the rotational axis. In rotating RBC, acceleration 
fluctuations in the horizontal direction are found to be enhanced by rotation at the transition from the 
rotation-unaffected to the rotation-affected regime, see Ref. \cite{Rajaei2016}. In that paper \cite{Rajaei2016}, it was shown 
that this transition towards a more intermittent acceleration is related to the transition in the BLs from 
the Prandtl--Blasius type to the Ekman type. In particular, the spiraling motion of plumes developing in 
the Ekman BL enhances the horizontal component of the acceleration. Within the resolution of the Rossby number used 
in \cite{Rajaei2016}, a gradual transition in the Lagrangian statistics of acceleration was observed. The 
gradual change of the Lagrangian statistics has not been explored in more detail in that particular study, 
although it was somewhat unexpected, given that the (related) transition in the heat transfer is known to be sharp.\\
\indent Here, we will extend the experimental study of Ref. \cite{Rajaei2016} numerically by zooming in on the transition 
in the Lagrangian acceleration statistics in order to better understand the flow dynamics around 
this transition and the relation between the Lagrangian velocity and acceleration of fluid particles and the thermal plumes. We collect 
these Lagrangian velocity and acceleration statistics in rotating RBC using direct numerical simulations (DNS) over a considerable 
range of rotation rates, between $Ro=0.058$ and $Ro=\infty$ (no rotation). In particular, we enhance 
the resolution of 
measurement points (Rossby numbers) around the anticipated transition at $Ro_c$ observed in the heat transfer, which from now on will be 
denoted as $Ro_{c_1}$ as maybe a second critical Rossby number may occur. Different regions of the flow are explored 
and fluid particle velocity, acceleration and vorticity statistics measured in the center of the convection cell are 
compared to similar statistical quantities measured near the top plate (including the BL, but with a 
substantial measurement region outside the BL). The advantage of using DNS is that the flow field (an Eulerian 
dataset) and the fluid particle trajectories (a Lagrangian dataset) are available simultaneously, so that 
we can also make a direct connection between the Lagrangian velocity and acceleration statistics and the characteristics of the 
underlying flow field such as the (vertical) vorticity. In this work, the focus is on finding the critical Rossby number for 
the transition in the dominant flow structures (which we denote here by $Ro_{c_2}$ as we cannot assume a priori that this critical Rossby number is the same as the one 
mentioned above and denoted by $Ro_{c_1}$) and the possible abruptness of such a transition in terms of the 
critical Rossby number $Ro_{c_2}$, and it is based on {\it{Lagrangian statistics data}}, in particular the autocorrelation 
fluctuations and higher-order statistics. Additionally, we report signatures of the transition in the flow 
structures close to the plates in rotating RBC.\\
\indent The paper is organized as follows: In Section II we introduce the numerical method used in this investigation. Results with regard to the 
transition in the heat transfer and the acceleration statistics, and transitions in the flow structures 
near the top plate, are presented and discussed in Section III. Section IV contains a summary and the 
main conclusions.
\section{Numerical method}
\begin{figure}
\includegraphics[width=0.4\textwidth]{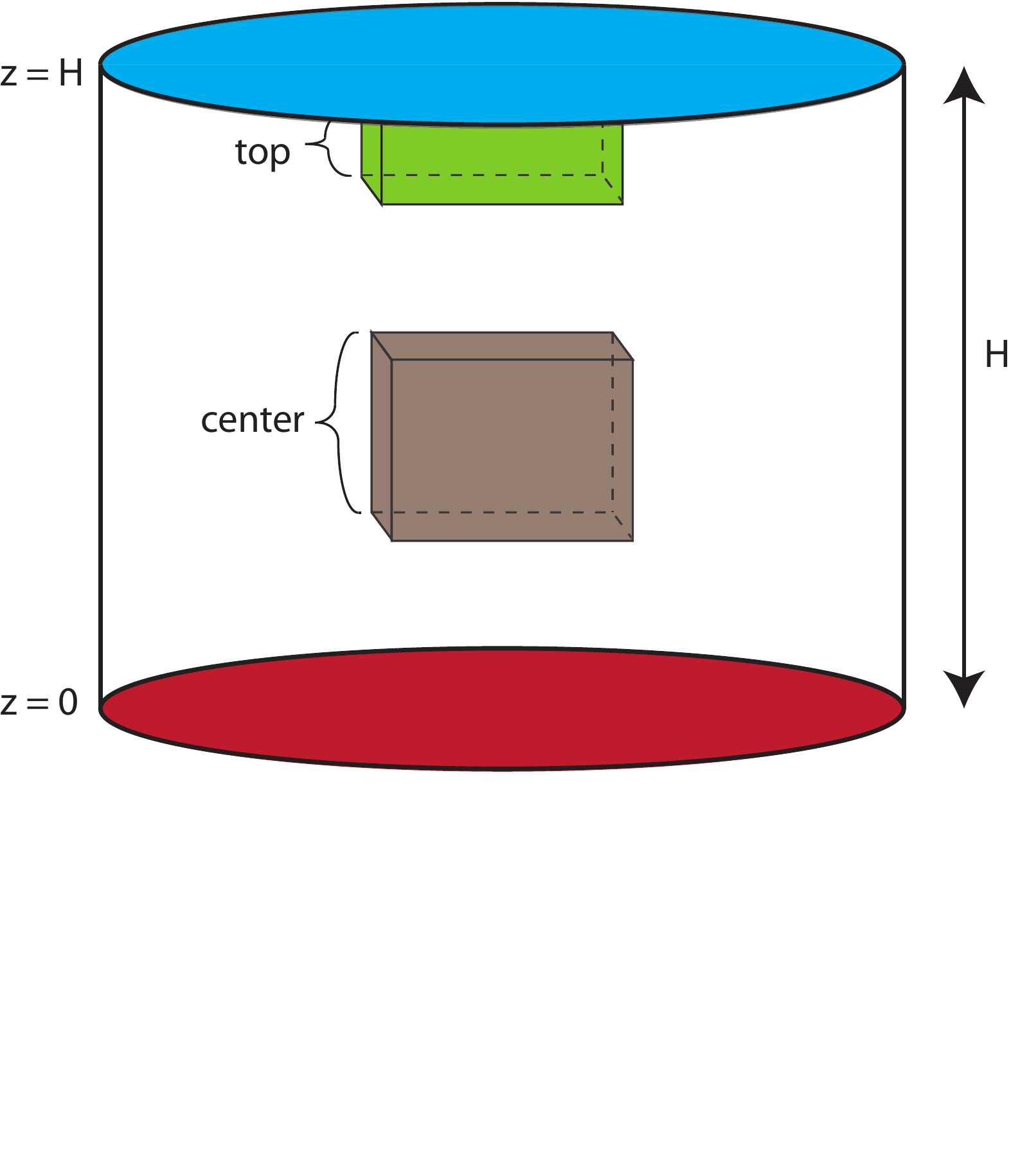}
\caption{\raggedright Sketch of the measurement volumes (not to scale). The gray cube in the center represents a 
measurement volume of size $0.25H \times 0.25H \times 0.25 H$, in the $x-$, $y-$ and $z-$direction, 
respectively. The green rectangular parallelepiped at the top plate represents a measurement volume of 
size $0.25H \times 0.25H \times 0.05H$ that starts right under the top plate (spanning the range $0.95H < z < H$ vertically).}
\label{fig:volumes}
\end{figure}

We have performed DNS of a (rotating) cylindrical Rayleigh--B\'{e}nard system. The governing dimensionless 
equations are the incompressible Navier--Stokes equations with the Coriolis term, and the energy equation, 
both in the Boussinesq approximation: 
\begin{align}
	{\boldsymbol{\nabla}} \cdot \textbf{u} &= 0 \label{eq:cont}, \\
	\frac{\partial \textbf{u}}{\partial t} + (\textbf{u}\cdot \pmb{\nabla}) \textbf{u} + \frac{1}{Ro} \hat{\textbf{z}} \times \textbf{u} &= - \pmb{\nabla} p + \sqrt{\frac{Pr}{Ra}} {\boldsymbol{\nabla}}^2 \textbf{u} + T \hat{\textbf{z}}, \label{eq:NS} \\
	\frac{\partial T}{\partial t} + (\textbf{u} \cdot \pmb{\nabla}) T &= \frac{1}{\sqrt{Pr Ra}} \nabla^2 T, \label{eq:energy}
\end{align}
with $\textbf{u}$ the velocity vector, $t$ time, $p$ pressure, $T$ temperature and $\hat{\textbf{z}}$ the 
vertical unit vector. These equations are non-dimensionalized using the cell height $H$ for length, 
$\Delta T$ (the temperature difference between the bottom and top plate) for temperature, and $t_c=H/U$ for time, based on the free-fall velocity 
$U \equiv \sqrt{g \alpha \Delta T H}$, where $g$ is the gravitational acceleration and $\alpha$ 
is the thermal expansion coefficient of the fluid. The corresponding dimensionless numbers are the Rayleigh number 
$Ra = g \alpha \Delta T H^3 / (\nu \kappa)$, the Prandtl number $Pr = \nu / \kappa$, and the Rossby number 
$Ro = U / (2 \Omega H)$, with $\nu$ and $\kappa$ the kinematic viscosity and thermal diffusivity of the 
fluid, respectively, and $\Omega$ the rotation rate. We simulate a cylinder with aspect ratio 
$\Gamma = D / H = 1$, with $D$ the diameter of the cell. We solve the equations in 
cylindrical coordinates using a second-order finite difference scheme that is described in detail 
in \cite{Verzicco2003,Verzicco1996}. For the discretization \num{512 x 384 x 512} grid points are used in 
the azimuthal, radial, and axial direction, respectively. To ensure that there are at least ten grid points within 
the boundary layer, grid refinement towards the walls is used in both the vertical and radial directions. The boundary 
conditions (BCs) are no-slip BCs at all walls, a fixed temperature BC at the top/bottom horizontal walls 
and adiabatic BCs (i.e. absence of heat flux) at the sidewalls. The other control parameters are set as 
$Ra = 1.3 \times 10^9$, $Pr = 6.7$ (corresponding to water) and the Rossby number is varied between 
$0.058 \leq Ro \leq \infty$, where $Ro = \infty$ is the non-rotating case.\\
\indent Inside the RBC flow passive tracers, following the fluid motion exactly, are tracked. To interpolate the 
fluid velocity from the surrounding eight grid points around the particle position, we use a tri-linear 
interpolation scheme and for the time integration a second-order Adams--Bashforth scheme is used. 
Lagrangian acceleration statistics of $10^6$ passive tracers are collected in two different measurement 
volumes; one of size $0.25 H \times 0.25 H \times 0.25 H$ placed in the center of the cell and one of 
size $0.25 H \times 0.25 H \times 0.05H$ placed under the top plate as sketched in figure \ref{fig:volumes}. 
This top measurement volume is attached to the top plate such that it spans the range $0.95H < z < H$ 
vertically, while it is centered around $r = 0$ horizontally. In order to obtain error estimates we use the symmetry of the 
flow problem with respect to the plane $z=0.5H$ and compare statistical data obtained for $z>0.5H$ with those obtained for 
$z<0.5H$. For the symmetric domain in the center this can be implemented in a straightforward way. 
We also analysed Lagrangian acceleration statistics 
of passive tracers in a similar domain near the bottom plate, spanning the vertical range $0 < z < 0.05H$, to complement 
the data from the volume just below the top plate. We will in addition collect more local 
statistics in horizontal slabs of thickness $\Delta z_i=0.001H$ and different vertical positions $z_i$. 
\section{Results and discussion}
\subsection{Transition in the heat transfer}
\begin{figure}
\centering
\includegraphics[width=0.48\textwidth]{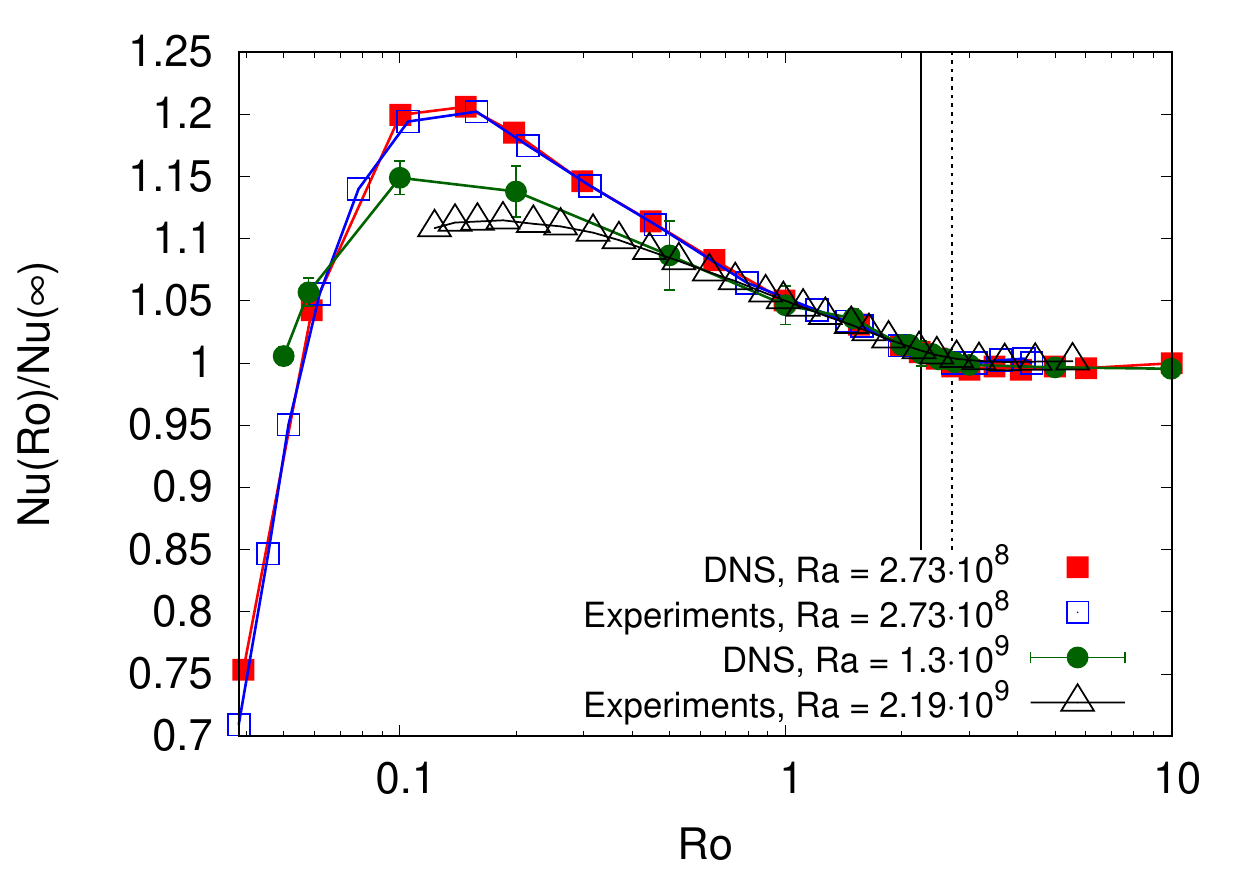}
\includegraphics[width=0.48\textwidth]{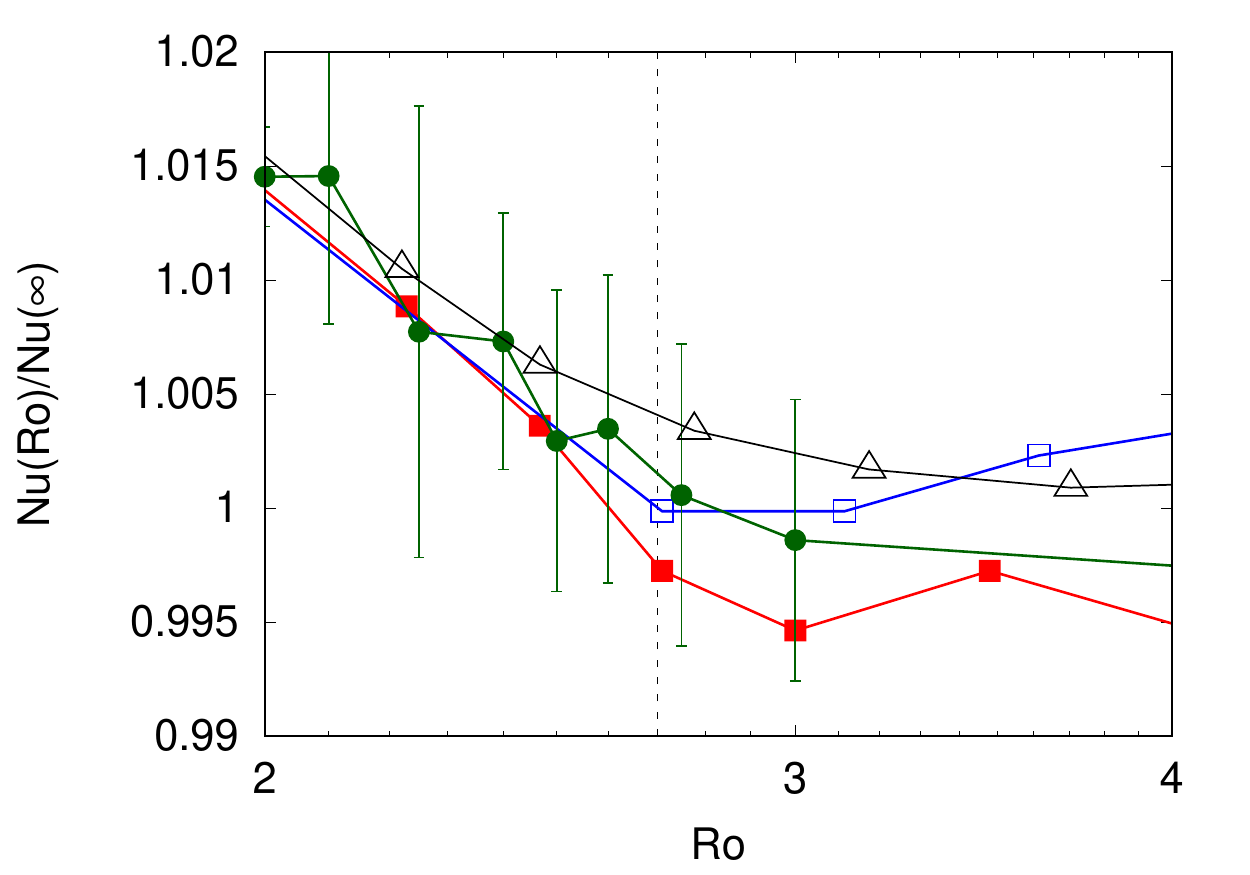}
\caption{a) The Nusselt number $Nu$ as function of $Ro$, normalized by $Nu(\infty)$ for the non-rotating 
case. Squares show data from \cite{Stevens2009,Zhong2009} for $Ra = 2.73 \times 10^8$ and 
$Pr = 6.26$, circles show data for the current simulations at $Ra = 1.3 \times 10^9$ and $Pr = 6.7$, and triangles 
show experimental data by \cite{Zhong2010} (supplementary data) taken at $Ra=2.19\times 10^9$ and $Pr=6.26$. 
Closed symbols are for DNS while the open symbols are for experiments. b) The same datasets, but now 
showing a zoom of the data in the range $2 \le Ro \le 4$. The vertical dotted line in both panels 
represents $Ro_{c_1} = 2.7$, and the vertical solid line in the left panel represents $Ro_{c_2}=2.25$.}
\label{fig:Nu}
\end{figure}

For the set of parameters studied here, it is known that there is a sharp transition in the heat 
transfer around a critical Rossby number $Ro_{c}$ in the range $2.3\lesssim Ro\lesssim 2.9$ \cite{Kunnen2008,Stevens2010a,Wei2015,Zhong2009}, see also Table \ref{Tab1}. 
The heat transfer is expressed by the Nusselt number $Nu$, giving the ratio between the total heat transfer 
and the convective heat transfer. In (rotating) RBC, the Nusselt number averaged over the full volume can be 
written as $Nu = 1 + \sqrt{Pr Ra} \langle u_z T \rangle$, where $u_z$ is the vertical fluid velocity. In 
figure \ref{fig:Nu}, we show this Nusselt number as a function of $Ro$ from the current simulations 
and from data reported in \cite{Stevens2009,Zhong2009,Zhong2010}. We observe a transition from the constant heat 
transfer regime (where $Nu(Ro)/Nu(\infty) \approx 1$) to an enhanced heat transfer regime at 
$Ro_{c_1} \approx 2.7$, 
presented by the dotted line (here, we introduce the critical Rossby number $Ro_{c_1}$ for the specific case 
explored in this study). Note that for the current discussion there is no need to pinpoint the 
exact position of the transition.
\subsection{Acceleration statistics}
As mentioned in the Introduction, we will supplement the work of Rajaei {\it{et al.}}~\cite{Rajaei2016}, where 
trajectories of neutrally buoyant particles are reconstructed in experiments of rotating RBC with 
$Ra = 1.3 \times 10 ^9$ and $Pr = 6.7$, by numerical simulations. In these numerical simulations the resolution 
in the Rossby number around $Ro_{c_1}$ is increased compared to the investigation reported in Ref.~\cite{Rajaei2016} 
to have a closer look at the position of the 
transition in terms of the Lagrangian acceleration statistics (and possibly disclosing its sharpness). 
Given that the flow structures in and nearby the BLs at the horizontal plates are expected to drive the 
transition, we will compare Lagrangian statistics collected both in the center of the convection cell and 
in the top measurement volume as in \cite{Rajaei2016}. To understand how the fluctuations in the vertical 
and horizontal components of the acceleration change under rotation, we will first focus on the 
root-mean-square (rms) values of the horizontal and vertical acceleration. Then the kurtosis and skewness 
measurements are discussed. The kurtosis of the acceleration probability density function (PDF) gives an indication 
of the presence of extreme acceleration events and with the skewness of the acceleration PDF 
we can explore whether the Lagrangian acceleration statistics are symmetric or asymmetric. 
\subsubsection{Root-mean-square values of acceleration}
\begin{figure}
\subfloat[]{\includegraphics[width=0.5\textwidth]{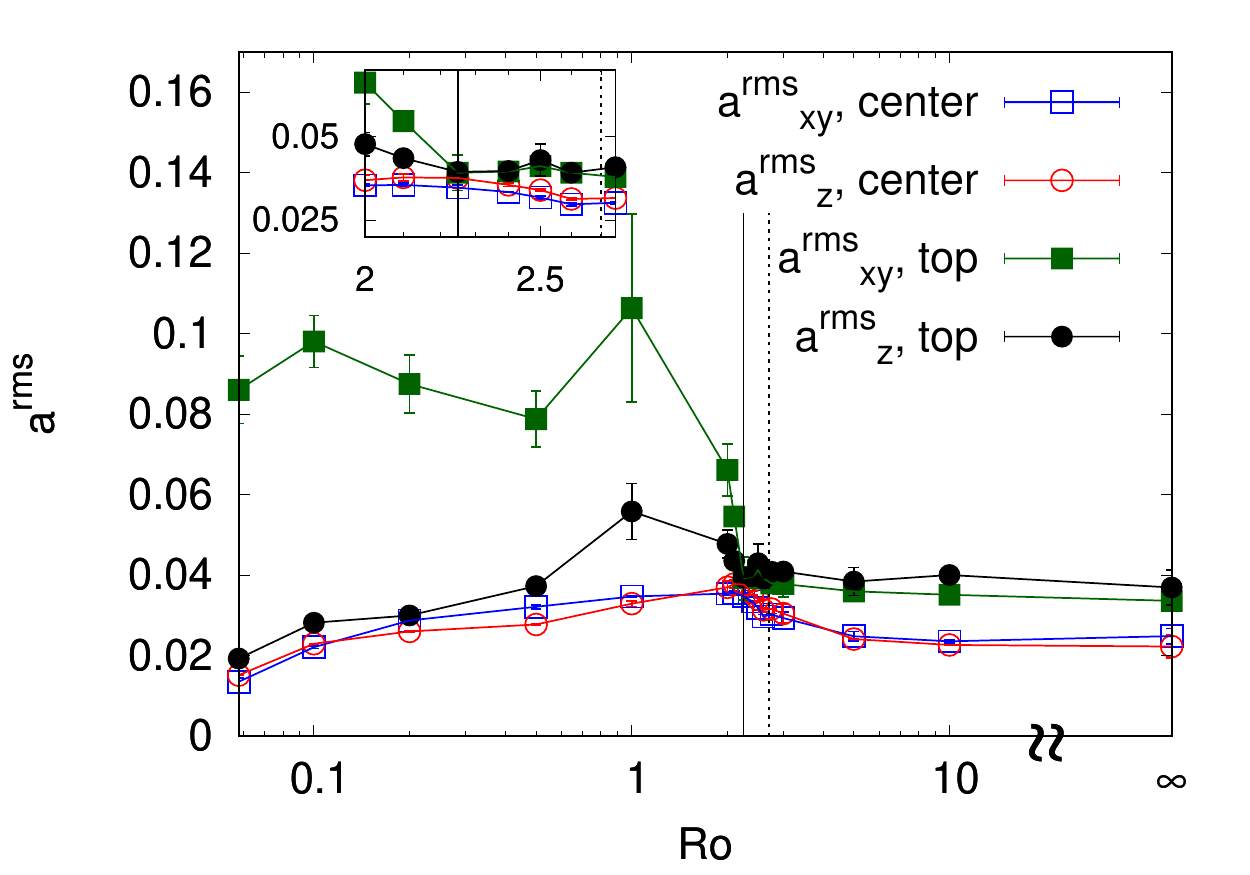} \label{fig:arms}}
\subfloat[]{\includegraphics[width=0.5\textwidth]{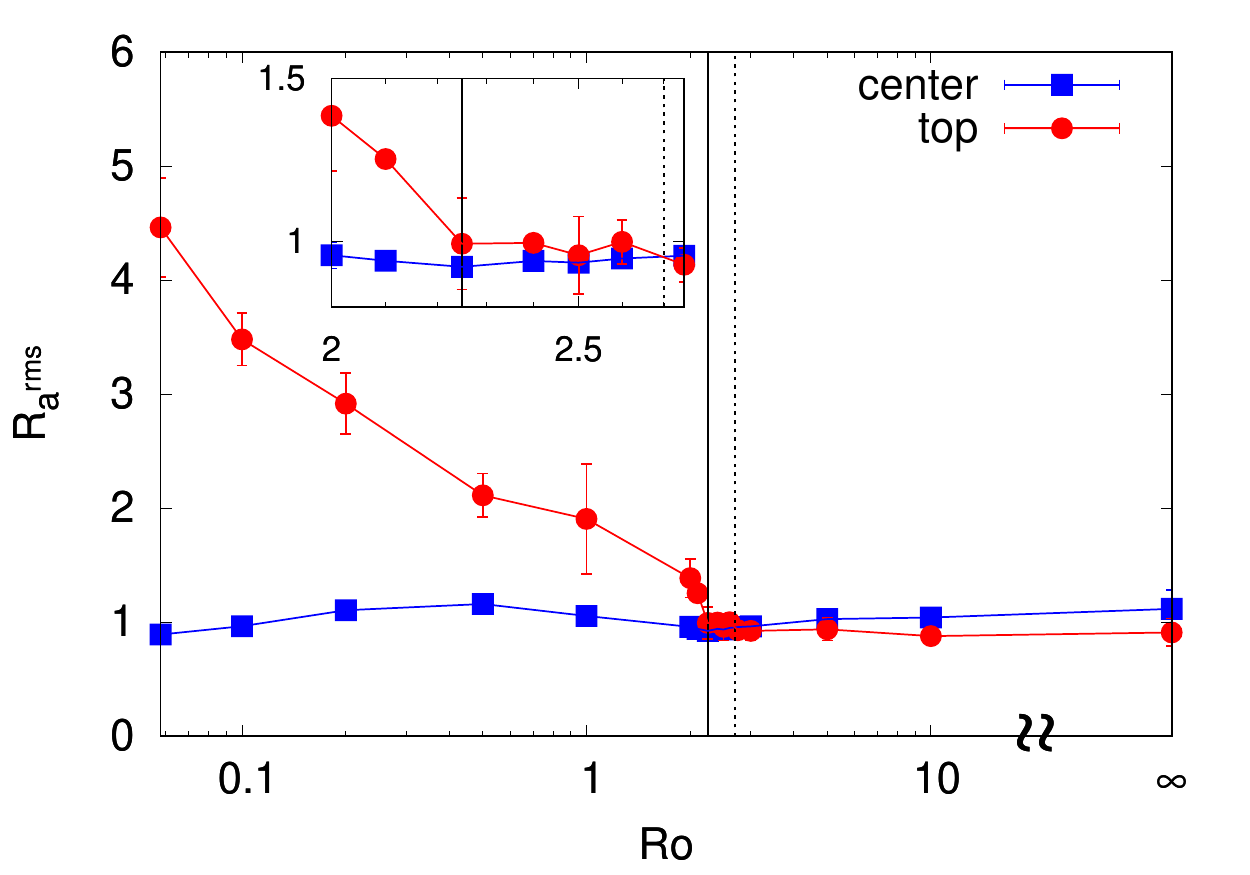} \label{fig:ratioarms}}
\caption{(a) Root-mean-square (rms) values of the horizontal and vertical acceleration components, 
$a^{rms}_{xy}$ and $a^{rms}_{z}$, respectively, as a function of $Ro$. Statistics are collected in the 
center (open symbols) and near the top plate (closed symbols). (b) The ratio between the rms values of the 
horizontal and vertical acceleration components, $R_{a^{rms}} = a^{rms}_{xy}/ a^{rms}_{z}$, for the center 
(blue crosses) and the top (red asterisks). In both panels the inset shows a zoom of the data in the range 
$2 \le Ro \le 2.75$ and the vertical solid line represents $Ro_{c_2}=2.25$, the second critical 
Rossby number, clearly different from $Ro_{c_1}=2.7$ (shown as the dotted line). Some of the error bars have the same size as the 
symbols, or are even smaller, and therefore not visible (the error bars are based on the differences between the results 
from the Lagrangian data obtained in the top and bottom half of the convection cell).}
\label{fig:rms}
\end{figure}

The rms values of the horizontal and vertical acceleration are computed as $a^{rms}_i = \sqrt{ \langle (a_i - \langle a_i \rangle)^2 \rangle }$, 
where $i = xy$ or $i = z$ and the average is taken over time and over the top and center measurement volumes, respectively, 
as sketched in figure \ref{fig:volumes}. For the horizontal component, $a^{rms}_{xy}$, the average is taken 
over a statistical sample that includes the values of both $a_x$ and $a_y$ because the statistical 
properties of the turbulent flow are assumed to be homogeneous and isotropic in the horizontal direction 
(and cannot depend on the orientation of the horizontal coordinate axes). As already discussed in \cite{Rajaei2016} and shown by previous 
experiments on rotating turbulence \cite{Castello2011}, rotation decreases the acceleration intensity along 
the rotational axis. Indeed, $a^{rms}_z$ is found to decrease with decreasing $Ro$ both in the center and 
near the top plate when $Ro \lesssim 1.0$. However, and in contrast to \cite{Castello2011}, weak maxima in 
$a^{rms}_z$ are observed in rotating RBC which are positioned at $Ro \approx 2$ and 
$Ro \approx 1$ for the central and top measurement volumes, respectively. Note that these maxima do not 
coincide with the critical Rossby number $Ro_{c_1}\approx 2.7$ at which we observe the transition in heat transfer. 
With increasing rotation rate (decreasing $Ro$), the rms of the horizontal acceleration component $a^{rms}_{xy}$ 
in the center of the convection cell first slightly increases up to $Ro\approx 1$ and then decreases 
subsequently. This trend is opposite to what 
has been reported for rotating turbulence by \cite{Castello2011}. Near the top plate 
$a^{rms}_{xy}$ increases significantly with increasing rotation rate, due to the formation of vortical 
plumes with swirling horizontal motion in the Ekman BLs. As observed in the inset of figure \ref{fig:arms}, 
this transition to increasing $a^{rms}_{xy}$ near the top plate is quite sudden and occurs at $Ro_{c_2} \approx 2.25$, 
where $Ro_{c_2}$ is represented by the vertical solid line. The fact that this transition is more prominent 
for the rms values of horizontal accelerations points at an increase of the anisotropy of acceleration with 
increasing rotation for $Ro \lesssim 2.25$. We quantify this in terms of the ratio between the horizontal 
and vertical acceleration components, $R_{a^{rms}} = a^{rms}_{xy} / a^{rms}_{z}$, shown in 
figure \ref{fig:ratioarms}. As already found in the experiments of Ref.~\cite{Rajaei2016}, in the center of the convection cell 
this ratio is almost independent of the rotation rate and shows, near the top plate, a transition from an 
approximately constant anisotropy ratio $R_{a^{rms}}=a^{rms}_{xy}/a^{rms}_{z} \approx 1$ in the weakly 
rotating cases ($Ro \gtrsim 2.25$) to a 
trend with significant increase of the anisotropy ratio $R_{a^{rms}}$ for $Ro\lesssim 2.25$. 
In the inset of figure \ref{fig:ratioarms}, we show that also this transition at $Ro_{c_2}$ is quite abrupt. 
\subsubsection{Kurtosis and skewness of acceleration}
The probability density functions (PDFs) of acceleration in turbulent flows are characterized by exponential 
tails \cite{Voth2002,Toschi2009}. In (rotating) RBC, the coherent flow structures are influencing the tails of the 
acceleration PDFs \cite{Rajaei2016}. The relative importance of the tails in such a PDF can be quantified by 
the kurtosis, $K_i = \langle (a_i - \langle a_i \rangle)^4 \rangle / \langle (a_i - \langle a_i \rangle)^2 \rangle^2$, 
where the average is taken over volume and time. Extreme acceleration events can be observed in our DNS. As 
the most extreme of these events are relatively rare, typically not more than 1-3 counts in the histogram 
calculation, these extreme points of the PDFs are obviously and unavoidably not well-converged. Moreover, these extreme events will 
highly influence the kurtosis and make it difficult to observe a clear trend with the rotation rate. In the 
computation of the kurtosis we therefore only include acceleration events that occur with a probability 
$P_{a_i} > 10^{-4}$ in the PDFs.

\begin{figure}
\subfloat[]{\includegraphics[width=0.5\textwidth]{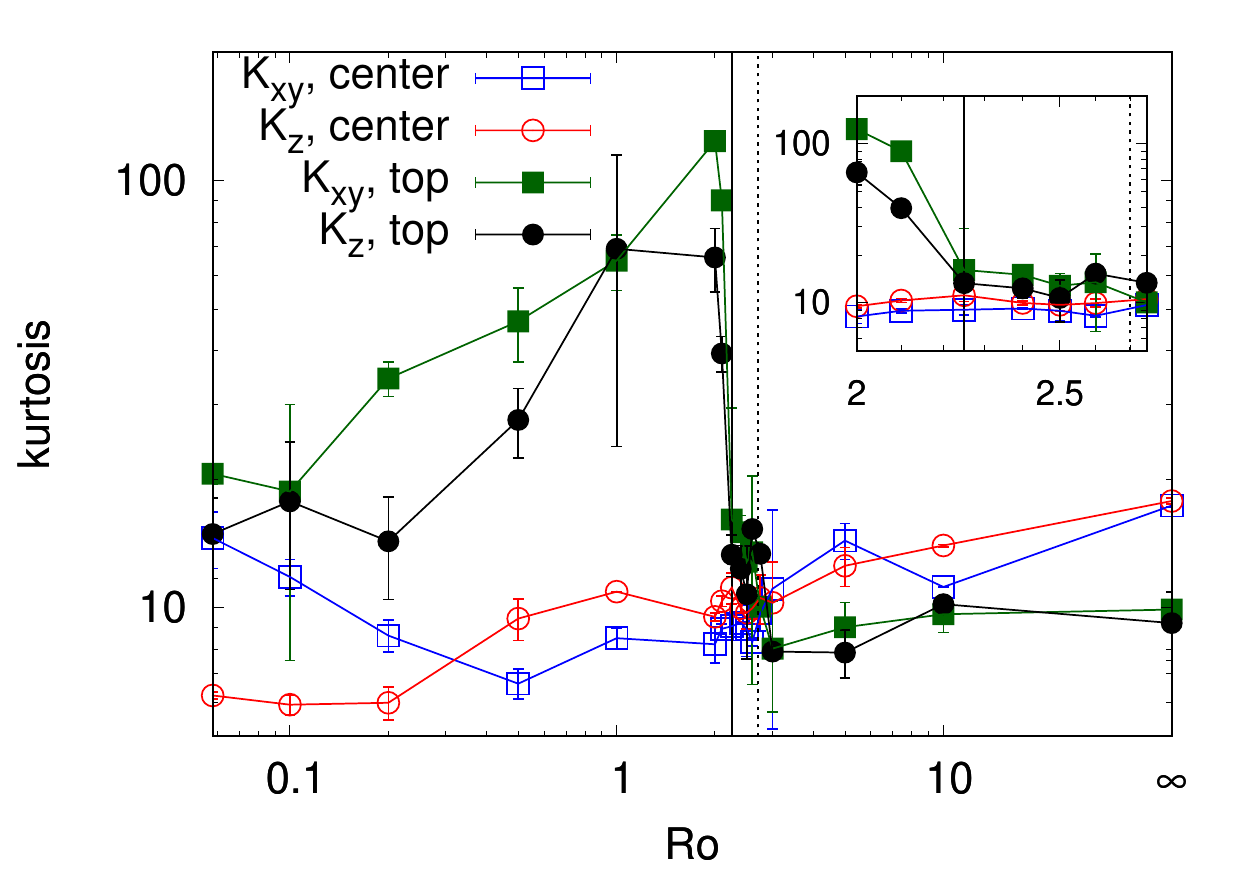} \label{fig:kurtosis}}~
\subfloat[]{\includegraphics[width=0.5\textwidth]{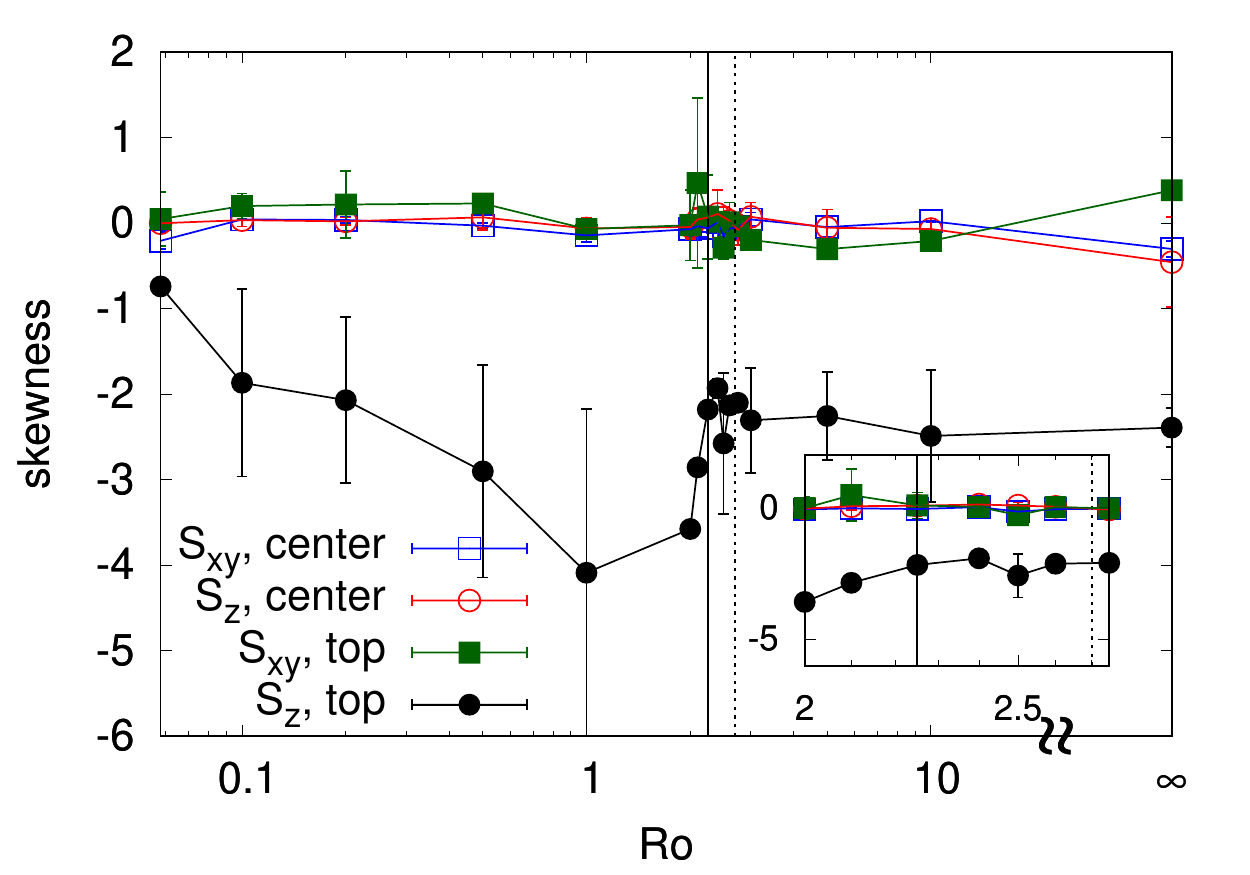} \label{fig:skewness}}
\caption{(a) Kurtosis and (b) skewness of the horizontal (squares) and vertical (circles) acceleration 
statistics as a function of $Ro$. Statistics are collected in the center (open symbols) and near the top 
plate (closed symbols). In both panels the inset shows a zoom of the data in the range $2 \le Ro \le 2.75$. 
The vertical solid line represents $Ro_{c_2}=2.25$, and the vertical dotted line $Ro_{c_1}=2.7$. Some of the error bars have the same size as the 
symbols, or are even smaller, and therefore not very well visible (the error bars are based on the differences between the results 
from the Lagrangian data obtained in the top and bottom half of the convection cell).}
\label{fig:kurtosis_skewness}
\end{figure}

In figure \ref{fig:kurtosis}, we show $K_{xy}$ and $K_z$ for both the center and top measurement volumes 
in the convection cell (for the location, again, see figure \ref{fig:volumes}). The values of the kurtosis measured in the DNS are 
in general more extreme than those measured 
experimentally in \cite{Rajaei2016}, where the difference can go up by a factor of four for both $K_z$ and 
$K_{xy}$ near the top plate and a factor of two in the center. This is a 
consequence of the extreme acceleration events in some (small-scale) vortex structures measured in the DNS 
which are washed out in the experimental data. Indeed, similar events are difficult to capture in the 
experiments where tracers always have a nonvanishing response time and where sudden movements of particles 
often results in loosing these particles in the experimental particle-tracking procedure. Although the 
values do not match one-to-one with those measured in 
\cite{Rajaei2016}, the trend with rotation in the kurtosis is quite similar. Like in \cite{Rajaei2016}, the kurtosis 
is only weakly affected by rotation in the center, where $K_{xy}$ has a local minimum around $Ro \approx 0.5$. 
This means that the PDFs of $a_{xy}$ are more intermittent for very small and very large Rossby numbers in the 
center, which is a result of the presence of coherent flow structures, which dominate the flow in the small 
and large $Ro-$number regimes (vertically-aligned vortices versus the LSC). Near the top plate both 
$K_{xy}$ and $K_{z}$ suddenly increase at $Ro_{c_2} \approx 2.25$ to reach their 
maximum value at $Ro \approx 2$, and decreasing again for $Ro \lesssim 2$. The quantitative values differ 
significantly from what has been found in the experiments by \cite{Rajaei2016}, although the trends as a 
function of $Ro$ are similar. With regard to these experimental data, this finding can be explained, on top of what has already been 
mentioned above with regard to the tracking algorithm, by the lack of (Lagrangian) data points 
inside the BL at the top plate in the experiments due to (i) a gap of about 1 \si{mm} 
between the top plate and the measurement volume and (ii) particles begin slightly heavier than the 
surrounding fluid (see also Ref.~\cite{Alards2017}). This BL is exactly the region where the thermal plumes develop that 
strongly accelerate the fluid horizontally (in the swirling plumes) and vertically away from the plate. In the 
experiments part of these extreme acceleration events 
near the top BL are thus missed, in particular the extreme vertical acceleration events in the 
swirling plumes. Since in the DNS the BL is fully included in the top measurement volume, 
extremer values of $a_z$ and $a_{xy}$ are included and hence extremer values of 
$K_z$ and $K_{xy}$ in the top measurement volume are obtained, compared to those from the experiments. All together, as in 
figure \ref{fig:rms}, the transition 
in the kurtosis observed near the top plate is very sudden and occurs at $Ro_{c_2} \approx 2.25$.\\
\indent The skewness, computed as $S_i = \langle (a_i - \langle a_i \rangle) ^3 \rangle / \langle (a_i - \langle a_i \rangle) ^2 \rangle^{3/2}$, 
gives a measure for the symmetry of the acceleration statistics. In figure \ref{fig:skewness}, we show 
$S_{xy}$ and $S_z$ as a function of $Ro$ for both the center and the top measurement volumes, where we only 
include acceleration events with a probability $P_{a_i} > 10^{-4}$ as we did for the kurtosis. In homogeneous isotropic 
turbulence (HIT), acceleration PDFs are symmetric and ideally $S_i = 0$. In the center, where the flow is closest 
to HIT \cite{Kunnen2010,Rajaei2016a,Zhou2008}, both $S_{xy}$ and $S_z$ indeed fluctuate around zero. Near 
the top plate, $S_{xy}$ fluctuates around zero, but $S_z < 0$, i.e., PDFs of $a_z$ are negatively 
skewed for all rotation rates. Plumes emerging from the BLs are accelerating the fluid moving away from 
the plates towards the bulk \cite{Julien1996}, resulting in more extreme negative acceleration events at 
the top plate. For $Ro \lesssim 2.5$ these plumes progressively become of the Ekman type and at 
$Ro_{c_2} \approx 2.25$, $S_z$ suddenly decreases (gets a more negative value, see figure \ref{fig:skewness}) to reach a 
minimum at $Ro \approx 1$ to then increases again (thus getting a less negative value) 
for $Ro \lesssim 1$, similar to the trend in the kurtosis that displays an extremum in the range 
$1 \lesssim Ro \lesssim 2$.\\
\indent The rms values, kurtosis and skewness of the Lagrangian acceleration statistics near the top plate all show a distinct 
transition at $Ro_{c_2} \approx 2.25$. This critical Rossby number is smaller than the Rossby number at 
which the behavior of the Nusselt number shows a clear transition, for our particular parameter settings at 
$Ro_{c_1}\approx 2.7$ as seen in figure \ref{fig:Nu}, thus 
$Ro_{c_2} < Ro_{c_1}$. By decreasing the Rossby number beyond $Ro_{c_2}$, the three quantities $a^{rms}$, $K$ and $S$ suddenly take very extreme values 
to then attenuate again for even lower $Ro$. This clearly suggests that there is also a sudden transition in the 
flow structures near the top plate at $Ro_{c_2} \approx 2.25$, which is possibly due to the transition from 
sheet-like thermal plumes in the Prandtl--Blasius type BL to vortical plumes emerging from the Ekman type BL. 
\subsection{Root-mean-square values of acceleration in horizontal slabs}
\begin{figure}
\includegraphics[width=0.5\textwidth]{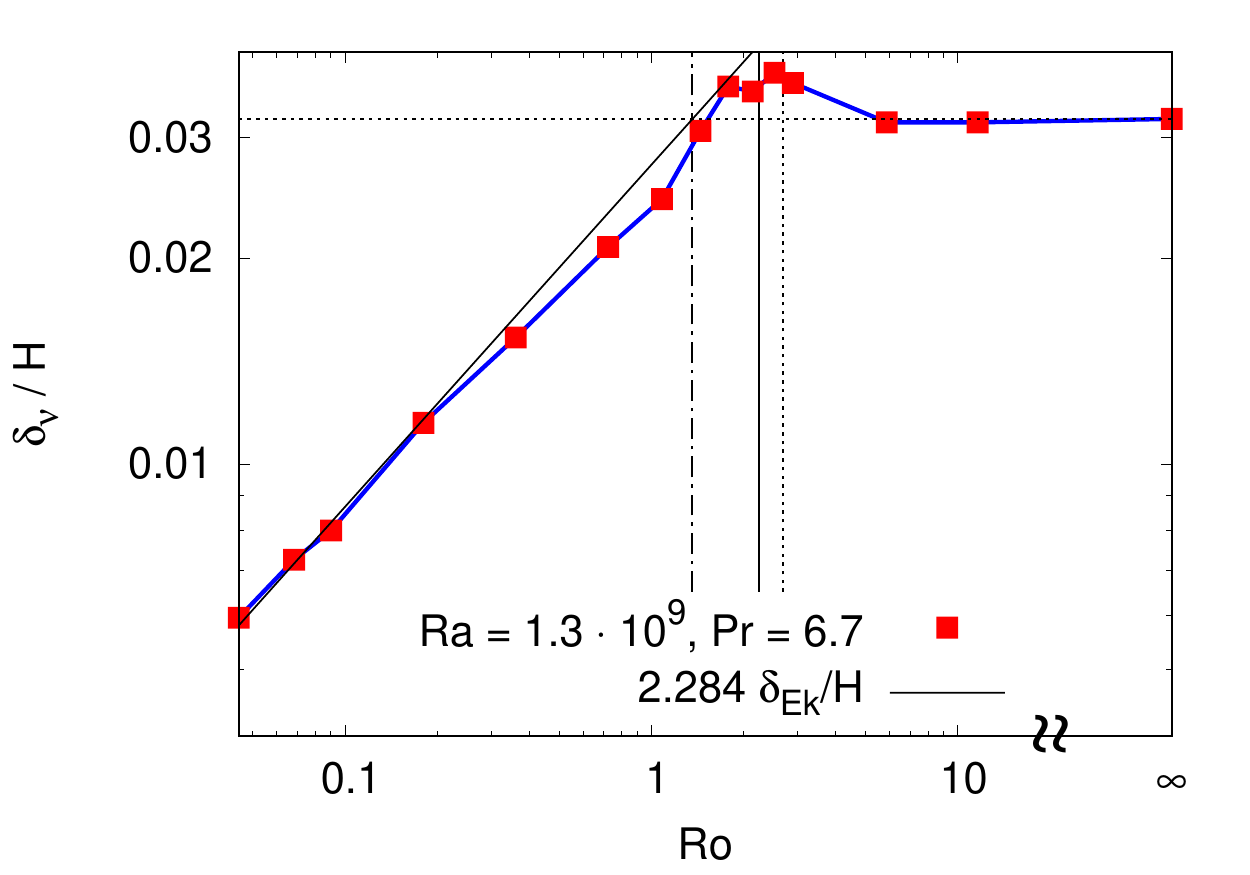}
\caption{The viscous BL thickness $\delta_{\nu}/H$ for cylindrical Rayleigh--B\'{e}nard 
convection with $Ra = 1.3 \times 10^9$, $Pr = 6.7$ and $\Gamma =1$. The BL thickness is measured from the 
DNS as the position of the maximum of the horizontal rms velocity. The horizontal dashed line shows the kinetic 
BL thickness $\delta_{\nu}/H=0.032$ and the sloping solid black line shows the theoretical prediction based on the linear Ekman BL theory, where 
$\delta_{Ek}=\sqrt{\nu/\Omega}$, see \cite{Rajaei2016}. The black dashed-dotted vertical line indicates the Rossby number 
where the viscous and Ekman BLs intersect ($Ro\approx 1.4$). The vertical solid and dotted lines indicate 
$Ro_{c_2}$ and $Ro_{c_1}$, respectively.}
\label{fig:BL}
\end{figure}

\begin{figure}
\subfloat[]{\includegraphics[width=0.5\textwidth]{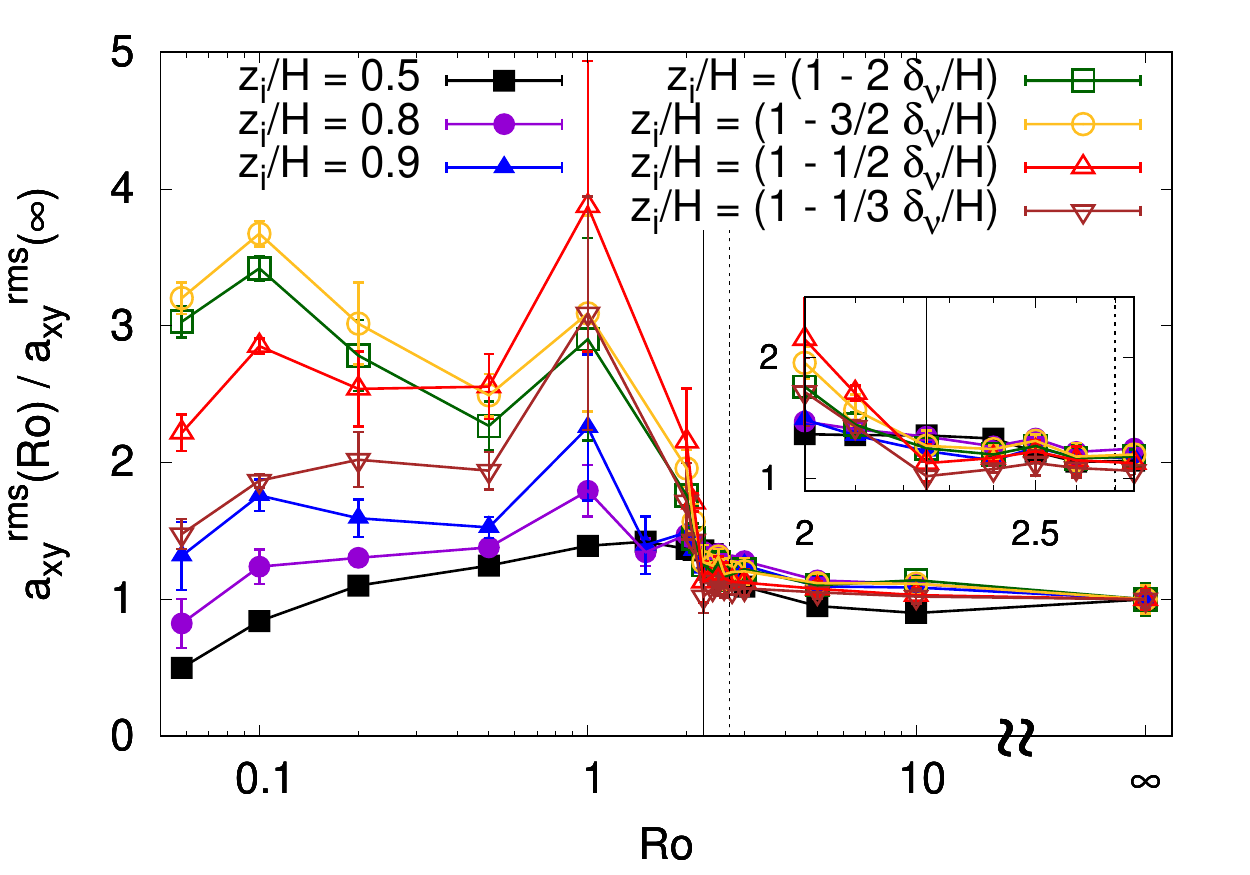} \label{fig:axyrms-zi} }~
\subfloat[]{\includegraphics[width=0.5\textwidth]{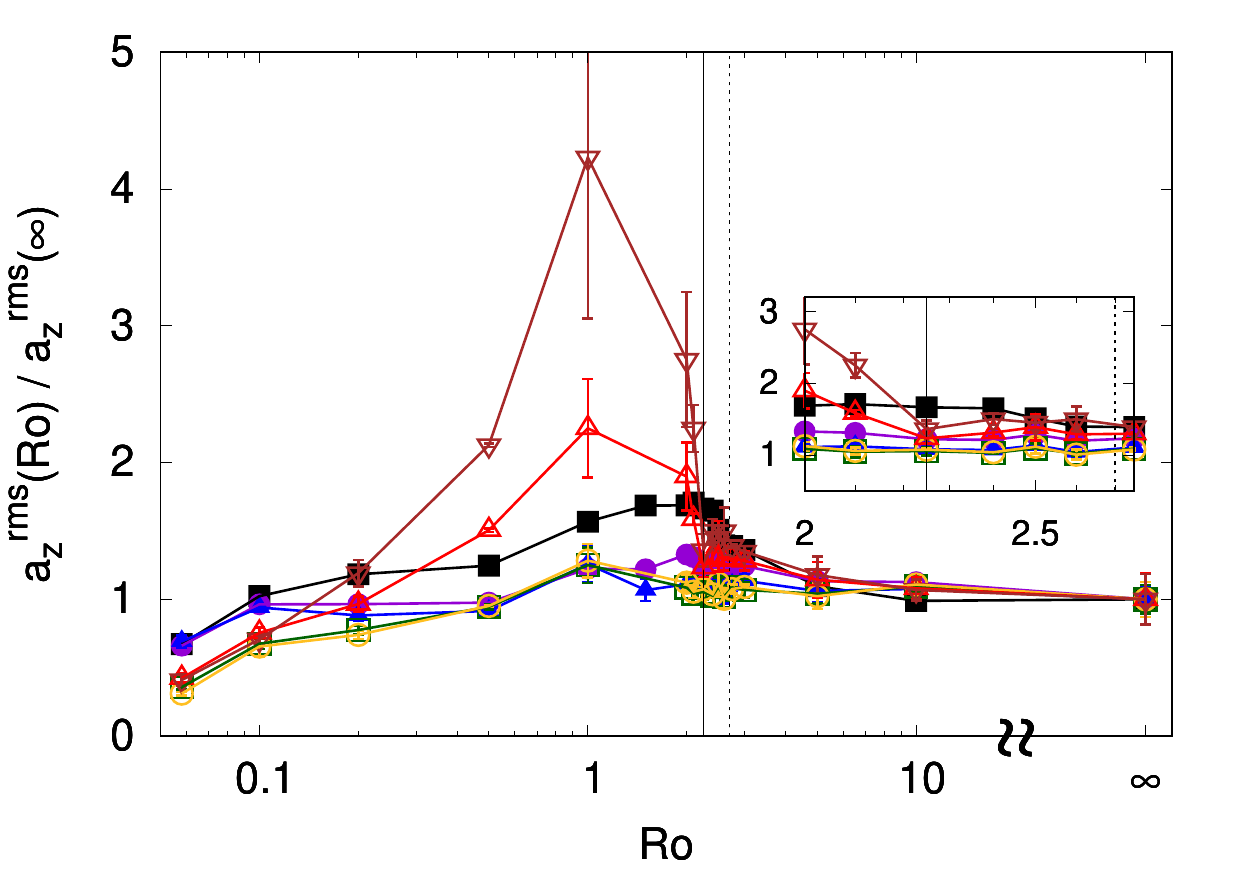} \label{fig:azrms-zi} }

\caption{Rms values of the (a) horizontal and (b) vertical acceleration components, both normalized by the 
respective rms acceleration values of the non-rotating case, $a^{rms}_{xy}(Ro)/a^{rms}_{xy}(\infty)$ and 
$a^{rms}_{z}(Ro)/a^{rms}_{z}(\infty)$, respectively. Rms values are computed in horizontal slabs of thickness 
$\Delta_{z_i} = 0.001H$ and central position $z_i$ as a function of $Ro$. The legend in panel (b) is the 
same as in panel (a) and in both panels the inset shows a zoom of the data in the range $2 \le Ro \le 2.75$. 
The vertical solid and dotted lines represent $Ro_{c_2}=2.25$ and $Ro_{c_1}=2.7$, respectively.}
\label{fig:arms-zi}
\end{figure}

So far, we have distinguished the center and top measurement volumes as sketched in figure \ref{fig:volumes}. 
We have found a sudden transition in the Lagrangian acceleration statistics at $Ro_{c_2} \approx 2.25$ near the top plate, while 
in the center such a transition is absent. Since the measurement volume near the top plate spans a height of 
$\Delta z = 0.05H$, it is not clear at which $z-$position the signatures of the transition become visible. 
We are therefore also not able yet to pinpoint the physical mechanisms responsible for this transition. To 
investigate the transition more locally, $a^{rms}_i$ is computed in horizontal slabs of thickness 
$\Delta_{z_i} = 0.001H$ centered at different vertical positions $z_i$. In the bulk, statistics are computed 
for $z_i /H = 0.5$, $z_i /H = 0.8$ and $z_i /H = 0.9$ (this latter value is most likely representing data 
in the bulk-BL mixed region). Near the BLs we take into account that the viscous 
BL thickness $\delta_{\nu}$ varies with the rotation rate and we express $z_i$ in terms of $\delta_{\nu}$. 
In figure \ref{fig:BL} the normalized BL thickness $\delta_{\nu}/H$ is shown as a function of $Ro$, computed from the current 
DNS as the position of the maximum of the horizontal rms velocity. Acceleration statistics are now computed 
in slabs with thickness $\Delta_{z_i} = 0.001H$ centered around $z_i=H-n\delta_{\nu}$, with $n\in\{2,\frac{3}{2},\frac{1}{2},\frac{1}{3}\}$. 
The results for 
$a^{rms}_{xy}$ and $a^{rms}_{z}$ are normalized by its value for $Ro = \infty$ in the corresponding slab, 
such that all curves have value unity at $Ro = \infty$ as shown in figure \ref{fig:arms-zi}. We see that 
the data for both $a^{rms}_{xy}$ and $a^{rms}_{z}$ almost overlap when $Ro \gtrsim 2.25$, while at $Ro_{c_2} \approx 2.25$ 
the trend suddenly changes. Values of $a^{rms}_{xy}$ start to increase with increasing rotation rate when 
$z_i/H \gtrsim 0.9$ and $Ro \lesssim 2.25$, see figure \ref{fig:axyrms-zi}. For comparison, the viscous BL thickness 
for $Ro > Ro_{c_2}$ is approximately constant and equals $\delta_{\nu}/H \approx 0.032$ (see 
figure \ref{fig:BL}) such that the BL at the 
top plate start at $z \approx 0.968H$ in this regime. For $Ro < Ro_{c_2}$ the BL thickness decreases up to 
a value of $\delta_{\nu} \approx 0.006H$ when $Ro = 0.058$ (see figure \ref{fig:BL}) and the BL at the top plate starts at 
$z \approx 0.994H$ for this rotation. Since the transition in $a^{rms}_{xy}$ is already visible from 
$z_i/H = 0.9$ on (a vertical position that is outside the viscous BL for all rotation rates), we can argue 
that Ekman plumes with a swirling horizontal motion emerge when $Ro < Ro_{c_2}$ are also felt outside 
the BL (up to a distance of at least $4\delta_{\nu}$ from the top plate).\\
\indent While $a^{rms}_z$ is only showing a weak increase at $Ro_{c_2}$ in the top measurement volume, see 
figure \ref{fig:rms}, a much stronger increase is observed in figure \ref{fig:azrms-zi} for 
$z_i/H \gtrsim 1-\delta_{\nu}/H$ (brown and red curves in figure \ref{fig:azrms-zi}), i.e., when statistics are 
collected inside the viscous BL. This indicates that rotation enhances the fluctuations in the vertical 
acceleration components only inside the BL, while it does not significantly affect these fluctuations 
inside the bulk. Since the BL is not turbulent in the parameter regime simulated here, this is expected to be purely related to 
the emergence of vortical plumes in the Ekman BL, accelerating the fluid away from the plate.\\
\indent All together, figure \ref{fig:arms-zi} once again shows a sudden transition in the Lagrangian acceleration statistics 
at $Ro_{c_2} \approx 2.25$ close to the top plate, where the magnitude of the increase of the rms-values 
at this transition depends on $z_i$. 
\begin{figure}
\includegraphics[width=0.99\textwidth]{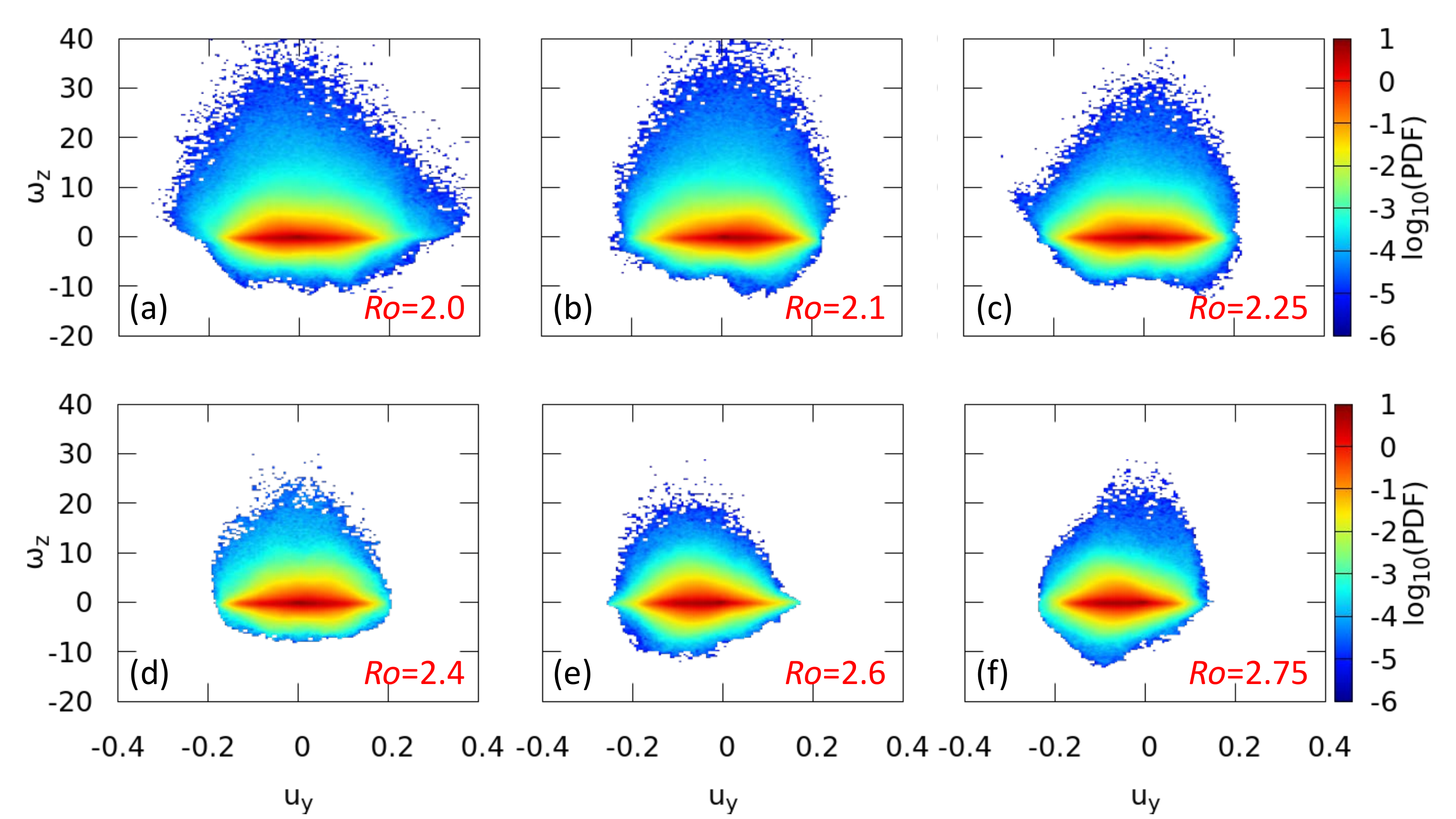}
\caption{Joint PDFs of the horizontal Lagrangian velocity of passive tracers, $u_{y}$, and the vertical vorticity, 
$\omega_z$, measured in the top measurement volume for six different Rossby numbers.}
\label{fig:jointPDfsuywz}
\end{figure}
\subsection{Flow structures near the top plate}
\begin{figure}
\includegraphics[width=0.99\textwidth]{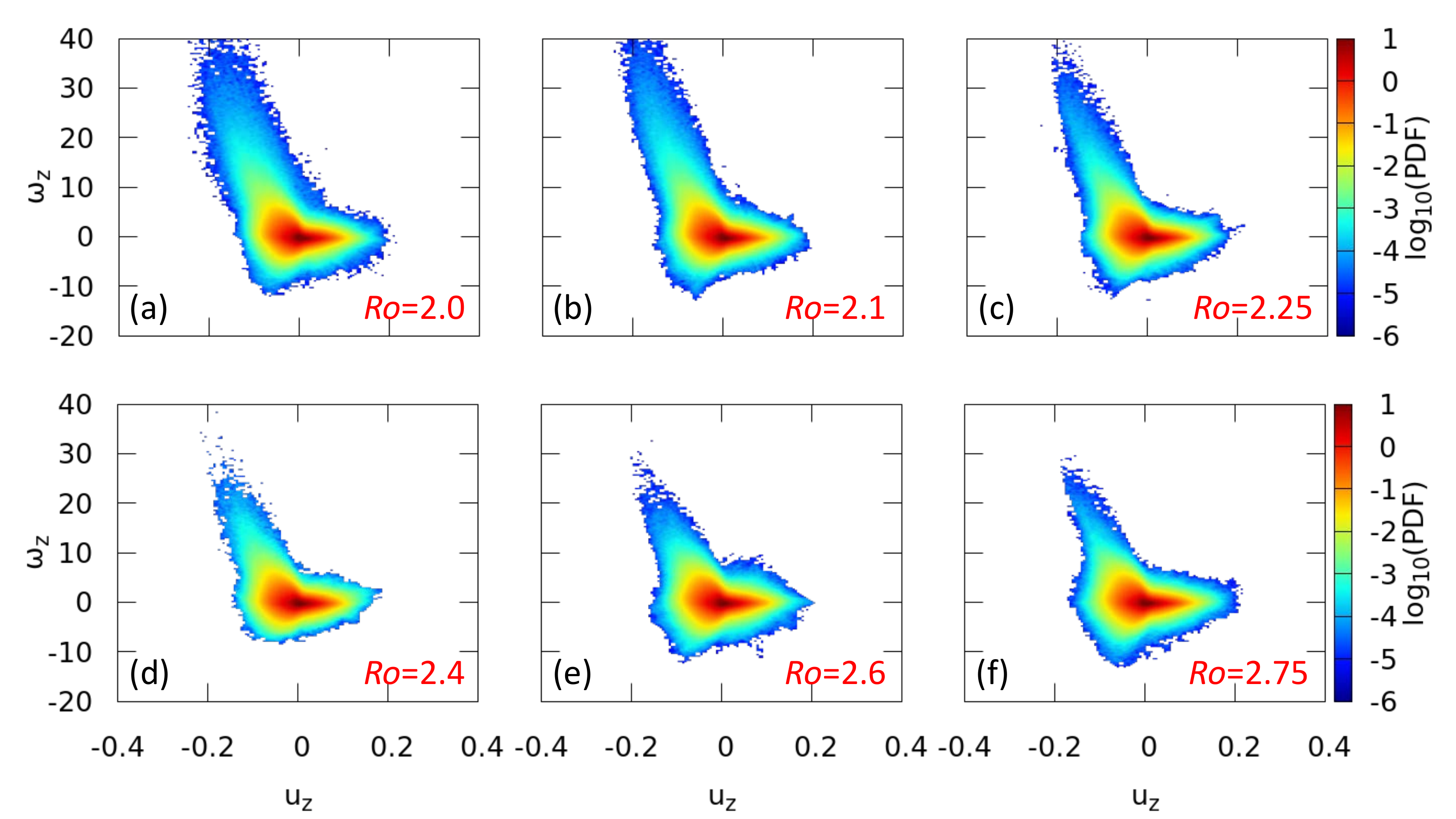}
\caption{Joint PDFs of the vertical Lagrangian velocity of passive tracers, $u_{z}$, and the vertical vorticity, 
$\omega_z$, measured in the top measurement volume for six different Rossby numbers.}
\label{fig:jointPDfsuzwz}
\end{figure}

To understand why the Lagrangian acceleration statistics show such an extreme transition at 
$Ro_{c_2} \approx 2.25$, and why it occurs at a smaller Rossby number than the transition in the 
behavior of the (normalized) Nusselt number ($Ro_{c_2}<Ro_{c_1}$), 
we need to better understand the underlying flow structures. For this purpose, 
we will consider joint PDFs of the vertical vorticity $\omega_z$ (obtained from the Eulerian 
flow field) with (i) the horizontal and vertical Lagrangian velocities $u_y$ and $u_z$, 
respectively, and (ii) the horizontal Lagrangian acceleration $a_{xy}$. Note that we do not use the horizontally-averaged 
$u_{xy}$, as we do for the acceleration statistics, because with the separate horizontal 
velocity components we can better distinguish the presence of the LSC. With the joint PDFs of 
vertical vorticity and 
horizontal velocity we can potentially provide some insight into the fate of the large-scale circulation (LSC) with 
decreasing $Ro$. On the other hand, the joint PDFs of vertical vorticity with the vertical component of 
velocity and horizontal components of acceleration are suitable to potentially visualize the 
formation of vertically-aligned vortical structures when approaching $Ro_{c_2}$. To improve the statistics we now use a slightly larger measurement volume of size 
 $0.4 H \times 0.4 H \times 0.25 H$, starting at $z=0.75H$ and ending at $z=H$, i.e., reaching all the way 
 to the top plate.\\
\indent In figure \ref{fig:jointPDfsuywz} we have plotted the joint PDFs of the horizontal Lagrangian velocity 
component, $u_y$, and the vertical vorticity $\omega_z$ for $Ro\in\{2.0, 2.1, 2.25, 2.4, 2.6, 2.75\}$. 
For the highest Rossby numbers we clearly see the remnants of the LSC as the maximum of the joint PDFs 
have negative $u_y$ and an almost symmetric vertical vorticity distribution (provided the joint probability $P_{joint}\gtrsim 10^{-3}$). A lack of symmetry in the vorticity distribution 
occurs for high positive vorticity values (with rather low probability) and is due to the emergence of 
weak cyclonic vortices near the top plate, while the (remnants of the) LSC are still dominant 
(see also discussion below). Similar joint PDFs 
with $u_x$ show a maximum with $u_x>0$. For $Ro=2.4$ the maximum of the joint PDF 
has moved to the origin $(u_y,\omega_z)=(0,0)$ and the vorticity distribution becomes asymmetric with a strong preference for 
positive vertical vorticity, indicating emergence of cyclonic vortices. These cyclonic vortices 
become stronger with decreasing $Ro$. For $Ro\lesssim 2.25$ the $\omega_z$-distribution 
becomes even more asymmetric and the $u_y$ are basically symmetrically distributed around $u_y=0$ (with predominantly 
positive $\omega_z$). This is a clear signature of the dominance of vertically-aligned cyclonic vortices 
which must possess the same amount of positive and negative $u_y$ (pure swirling flows in the vortex cores). 
The destruction of the LSC and 
formation of vertically-aligned cyclonic vortices is nicely confirmed by the joint PDFs of the 
vertical Lagrangian velocity component $u_z$ and the vertical vorticity $\omega_z$ for the same range of $Ro$, 
see figure \ref{fig:jointPDfsuzwz}. Particularly convincing is the 
strong correlation of positive vertical vorticity with negative vertical velocity for $Ro\lesssim 2.25$. 
Quite remarkably, a correlation between positive $\omega_z$ and negative $u_z$ is already visible for $Ro\gtrsim 2.4$ 
which suggests that weak cyclonic vortices are already being formed near the top plate while the LSC is being 
weakened (see figure \ref{fig:jointPDfsuywz}). These aspects can be confirmed and explored further with the 
joint PDFs of vertical vorticity and horizontal acceleration of passive tracers. 

It is expected that the thermal plumes developing in 
the BLs at the horizontal plates are responsible for the extreme horizontal and vertical acceleration of 
fluid parcels. For $Ro \gtrsim 2.25$ the results presented and discussed in the previous (sub)sections 
suggest that these should 
predominantly be sheet-like thermal plumes and the LSC (or large-scale remnants of it) induces a 
mean wind at the horizontal plates as confirmed by the joint PDFs of vertical vorticity and horizontal velocity above, 
deflecting the plumes into the direction of the mean horizontal flow \cite{Shishkina2013}. 
Thermal plumes in the Prandtl--Blasius BL are characterized by large values of both positive and negative 
vertical vorticity, while vortical plumes emerging from the Ekman BL are cyclonic, i.e. they spin up in the 
same direction as the applied rotation $\bm{\Omega}$, resulting in positive vertical vorticity 
\cite{Shishkina2008,Julien1996a}, with extreme values of $\omega_z$. 
In the regime unaffected or weakly affected by rotation the deflection of the plumes by the mean wind is 
expected to largely suppress and flush away emerging regions of extreme vertical vorticity. The LSC 
has become already much weaker for $Ro \lesssim Ro_{c_2}\approx 2.25$ \cite{Kunnen2008,Zhong2010,Stevens2013,Sterl2016} 
(see also figure \ref{fig:jointPDfsuywz} and Table \ref{Tab1}) and in this regime regions with large and extreme positive vertical vorticity are expected 
to be much more dominant and persistant. Therefore, in this rotation-affected regime we do expect the extreme (horizontal) 
acceleration of tracers to be much more clearly correlated to the swirling motion of vortical plumes. 

\begin{figure}
\includegraphics[width=0.99\textwidth]{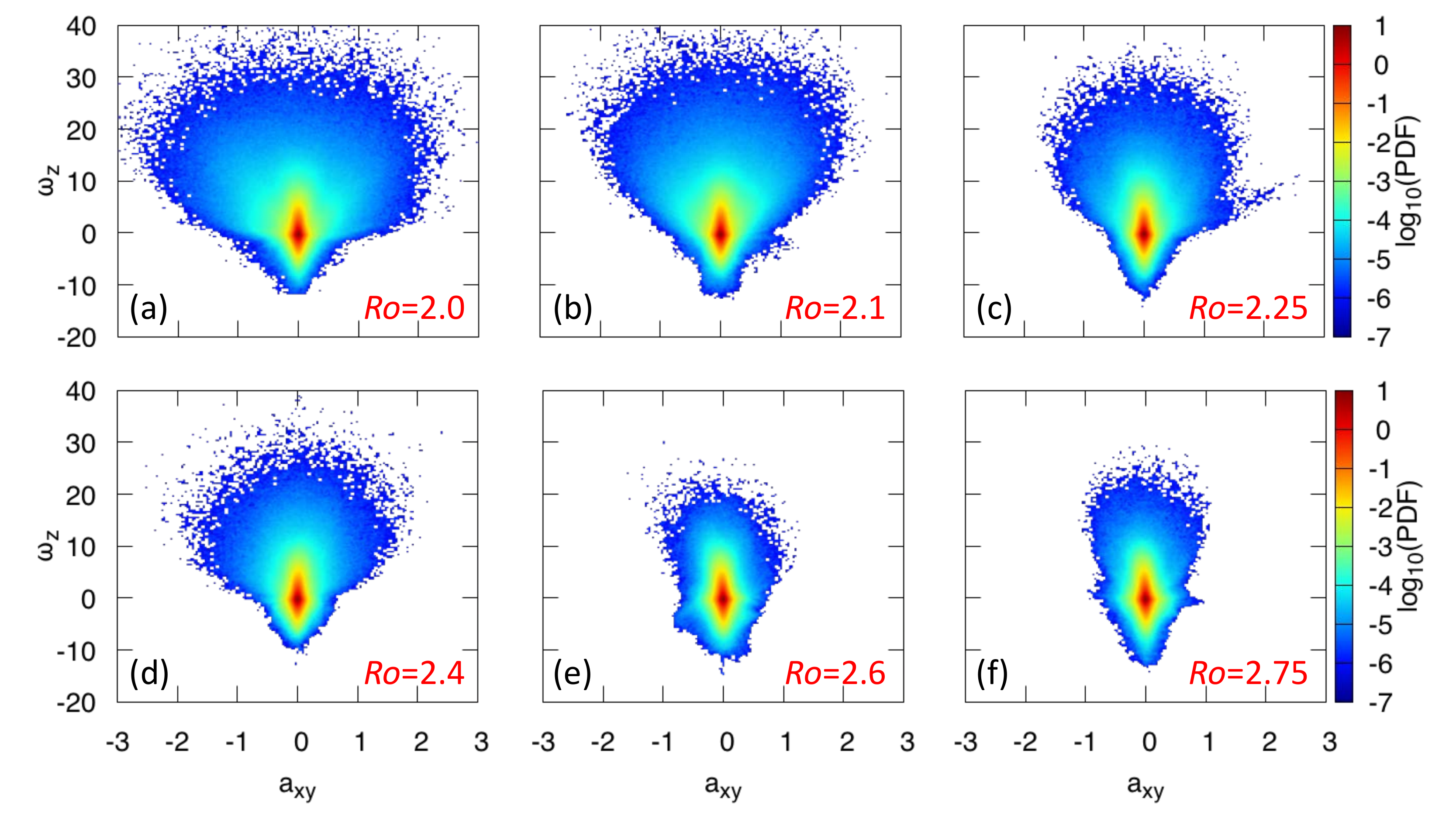}
\caption{Joint PDFs of the horizontal Lagrangian acceleration of passive tracers, $a_{xy}$, and the vertical vorticity, 
$\omega_z$, measured in the top measurement volume for six different Rossby numbers.}
\label{fig:jointPDfsaxwz}
\end{figure}

Vortical plumes in the Ekman BL spin up cyclonically and we thus expect extreme acceleration events to occur 
in regions with positive vertical vorticity. To investigate this relation between acceleration and vorticity, 
we compute the joint PDFs of the horizontal acceleration $a_{xy}$, and the vertical vorticity $\omega_z$. 
Results of these joint PDFs are shown in figure \ref{fig:jointPDfsaxwz} for six different Rossby numbers 
around the transition at $Ro_{c_2} \approx 2.25$. Although the behavior of the joint PDFs around this transition appears to 
be more gradual, a different behavior is clearly found before and after the transition at $Ro_{c_2}$. 
When $Ro \gtrsim 2.4$, the joint PDFs show an elongated patch centered in the origin ($\omega_z = 0$, $a_{xy}=0$), 
 and the most abundant acceleration events, that occur with a probability $P_{joint}>10^{-4}$, are almost 
 symmetrically distributed with respect to $\omega_z=0$ (see the area from red to light blue in the scatter 
 plots; the dark blue area contributes less strongly as there $10^{-7}\lesssim P_{joint}\lesssim 10^{-4}$), 
 indicating that acceleration events are correlated to small positive and negative values of $\omega_z$. 
 When $Ro \lesssim 2.4$, values of $\omega_z$ and $a_{xy}$ have 
 become more extreme (see the extension of the dark blue area) and now larger acceleration events, indicated by the red to light blue areas in the 
 joint PDFs (with the joint probability $P_{joint}> 10^{-4}$), are correlated to large values of mostly positive $\omega_z$. 
 The acceleration events that occur with a joint probability $P_{joint}>10^{-4}$ are thus asymmetrically 
 distributed around $\omega_z=0$. Although signatures 
 of the transition are already visible for $Ro \approx 2.4$, in particular for the extreme 
 acceleration events, the observations suggest that tracers exposed to large accelerations are trapped inside 
 Ekman plumes with cyclonic vorticity, most clearly for $Ro \lesssim Ro_{c_2}$.

\section{Conclusions}
We have investigated the effect of rotation on the Lagrangian acceleration statistics of passive tracers 
in RBC, where the focus is on the drastic change in the flow structures around the transition in heat transfer 
efficiency when going from the rotation-unaffected to the rotation-affected regime. The horizontal 
acceleration statistics near the top plate show a sudden transition at $Ro_{c_2} \approx 2.25$, 
below which the rms values and the kurtosis increase significantly. When collecting statistics in thin horizontal 
slabs at different vertical positions $z_i$, relative to the BL thickness, we find that the transition in 
the flow structures can be observed up to $z_i/H \gtrsim 0.9$, thus in the turbulent bulk quite far outside the 
boundary layer itself. Although rms values of vertical acceleration collected in the measurement volume near 
the top plate do not show a drastic 
transition at $Ro_{c_2}$, a sudden increase is found when measuring rms values of vertical acceleration 
inside the viscous BL. This is a Lagrangian signature of the presence of Ekman plumes strongly accelerating 
the fluid horizontally spiraling inwards into vertically-aligned vortices before being injected into the bulk. 
The kurtosis and skewness of $a_z$ do show a sharp transition in the top 
measurement volume.

Since transitions in the Lagrangian acceleration statistics are only found close to the top plate and not in the center of the cell, they are expected 
to be related to the development of thermal plumes in the BLs at the plates. Under rotation, these thermal 
plumes change from sheet-like plumes in the Prandtl--Blasius type BL regime to vertically-aligned vortical plumes in the 
Ekman type BL regime. To investigate the relation between the Lagrangian acceleration of fluid particles 
(or passive tracers) and the thermal plumes, we 
focused on the joint statistics of horizontal and vertical Lagrangian velocity components $u_y$ and $u_z$, respectively, with the vertical vorticity $\omega_z$ 
and on those of $a_{xy}$ and the vertical vorticity $\omega_z$. In the rotation-unaffected regime, joint PDFs of $a_{xy}$ and 
$\omega_z$ are symmetric around $\omega_z = 0$ and do not show significant extreme acceleration events 
when $Ro \gtrsim Ro_{c_2}$. This behavior completely changes when moving 
into the rotation-affected regime, in particular for $Ro \lesssim Ro_{c_2}$, where joint PDFs show a 
clear correlation between strong horizontal acceleration events and positive values of $\omega_z$. 
At the transition $Ro_{c_2}$, also the maximum values of $\omega_z$ increase drastically together with the 
emergence of extreme horizontal acceleration events. This suggests that the coherent structures 
in and near the Ekman BL are rather suddenly dominated by strong vertically-aligned vortical structures. 

This conclusion is supported by the joint PDFs of horizontal and vertical Lagrangian velocity with vertical vorticity, see 
figures \ref{fig:jointPDfsuywz} and \ref{fig:jointPDfsuzwz}. These PDFs indeed confirm the notion that the LSC 
does not disappear suddenly at $Ro_{c_1}$, but is slowly being destroyed. The remnants of the LSC are still able 
to sweep away emerging cyclonic vortices near the plates (which obviously tend to emerge 
for $Ro\gtrsim 2.4$, see figure \ref{fig:jointPDfsuzwz}). This is in agreement with observations 
in Ref. \cite{Kunnen2008} where it was concluded (with numerical simulations for $Ra=1.0\times 10^9$ 
and $Pr=6.4$) that the LSC becomes weaker for $Ro\lesssim 2.5$, but significant amounts of energy 
are still stored in the LSC motion down to $Ro\approx 1.2$, although rapidly decreasing 
with $Ro$. Experimental data on the properties of the LSC for $Ra=2.25\times 10^9$ and $Pr=4.38$ 
reveal a similar picture, see \cite{Zhong2010}. See also \cite{Stevens2013} where a 
different measure is applied for the LSC strength and an extended 
dataset from \cite{Zhong2010} is used for comparison. These experimental data clearly reveal that the LSC 
strength is significantly decreased when Rossby is reduced to 
$Ro\approx 1.2$. Also in Ref. \cite{Kunnen2010} it was observed that slightly enhanced values of horizontal and 
vertical root-mean-square velocities remain quite constant down to $Ro\approx 1.2$.

In previous experiments \cite{Rajaei2016} we were already able to use Lagrangian acceleration statistics 
to explore the transition. However, within the resolution of Rossby numbers used in that study the transition 
seemed gradual and, if present, it could still be connected with the known critical rotation rate for heat 
transfer enhancement (here indicated with $Ro_{c_1}$). In the present study we indeed see a sudden sharp 
transition in the Lagrangian acceleration statistics at $Ro_{c_2} \approx 2.25$. However, this critical 
Rossby number turns out to be {\it{smaller}} than $Ro_{c_1}\approx 2.7$. Thus at both rotation rates sharp transitions are 
observed, but for different quantities. {\COM{By looking at the acceleration of tracers and its relation to typical plume characteristics, 
we found that Ekman plumes are much more efficient in accelerating fluid parcels causing this sudden 
transition in the Lagrangian acceleration statistics.}} 

The most remarkable observation of this study is not the sharpness of the transition in the Lagrangian 
acceleration statistics, but the presence of two critical Rossby numbers while only one 
was anticipated in advance: (i) the well-known critical Rossby number $Ro_{c_1}$, indicating a sudden and sharp transition 
in heat transport properties in rotating RBC, and {\COM{(ii) a new critical Rossby number $Ro_{c_2}<Ro_{c_1}$ where 
the Lagrangian acceleration statistics show a sharp transition. The latter is intimately related with a change 
in dominant flow structures, i.e. from LSC to vertically-aligned vortices, which does not occur at $Ro_{c_1}$ but at $Ro_{c_2}$. Intuitively, one would 
expect that $Ro_{c_2}=Ro_{c_1}$ which is clearly not the case. The present observations hint at 
two BL related mechanisms responsible for these transitions and with which we can derive crude estimates for both 
Rossby numbers at which the transitions can take place.}}\\
In figure \ref{fig:BL} we have indicated the Rossby number at which the 
Ekman type BL and the viscous Prandtl-Blasius type BL have similar thickness. It was based on the assumption 
$2.284\delta_{Ek}=\delta_{\nu}$ (see the next paragraph for a brief motivation), or put slightly differently: 
$\delta_{Ek}/H = \sqrt{2Ro(Pr/Ra)^{1/2}} = \delta_{\nu}/(2.284 H)$. We can rewrite this as 
\begin{equation}
Ro=\frac{1}{2}\left(\frac{\delta_{\nu}}{2.284H}\right)^2\sqrt{\frac{Ra}{Pr}}~.
\label{eq:rossbyBL}
\end{equation}
The dimensionless BL thickness in our non-rotating RBC case (with $Ra=1.3\times 10^9$, $Pr=6.7$ and $\Gamma=1$) 
is $\delta_{\nu}/H\approx 0.032$, see figure \ref{fig:BL}. From Eq. (\ref{eq:rossbyBL}), the required Rossby 
number to satisfy $2.284\delta_{Ek}=\delta_{\nu}$ is then $Ro_{c_2,th}\approx 1.4$, 
somewhat smaller than $Ro_{c_2}\approx 2.25$, but consistent with Rossby-number estimates for the disappearance of 
the LSC \cite{Kunnen2008,Zhong2010,Stevens2013}.\\
\indent The first critical Rossby number, $Ro_{c_1}\approx 2.7$, reflects the rotation rate where 
$\frac{\partial {\bf{u}_h}}{\partial z}|_{z=0}$, with ${\bf{u}}_h=(u,v)$ in the Ekman type BL, 
becomes similar in magnitude to the normal gradient of the horizontal velocity component in the 
Prandtl--Blasius type BL at the wall. Note that under this condition the Ekman type BL thickness is then larger than the thickness 
of the Prandtl-Blasius type BL.\\
\indent To estimate the Rossby number for this to occur we compare the Ekman type BL 
with an approximation for the BL over the flat bottom and top plates in non-rotating RBC, the Blasius BL. 
The starting point is the fact that for similar thicknesses of these BLs the normal gradient at the 
plate is larger in magnitude for the Ekman BL compared to this gradient for the Blasius 
BL~\cite{Schlichting1960}. Using linear Ekman boundary-layer theory, see \cite{Kundu2002}, the 
matching of a geostrophic bulk flow with 
velocity $V$ (which we assume for convenience to be in the $x$-direction) to the solid wall gives 
the following expressions for the horizontal velocity field in the Ekman boundary layer: 
$u=V[1 - e^{-z/\delta_{Ek}}\cos (z/\delta_{Ek})]$ and $v=V e^{-z/\delta_{Ek}}\sin (z/\delta_{Ek})$. This immediately results in 
$\frac{\partial u}{\partial z}|_{z=0}=V/\delta_{Ek}$ and similarly for $\frac{\partial v}{\partial z}|_{z=0}$ 
(and the maximum of $\sqrt{u^2 + v^2}\approx 1.07V$ at $z=2.284\delta_{Ek}$, which we consider as 
the actual Ekman BL thickness and which is used as such in figure \ref{fig:BL}, before asymptotically 
reducing to $V$ for large $z$). Using the well-known properties of the Blasius BL that 
$\frac{\partial {\tilde{u}}}{\partial z}|_{{\tilde{z}}=0}=0.33$ and ${\tilde{\delta}}_{\nu}=4.9$ (in the 
non-dimensional BL units and the $u=0.99V$ criterion for 
$\delta_{\nu}$ \cite{Schlichting1960}), we find here that 
$\frac{\partial u}{\partial z}|_{z=0}=0.33\cdot 4.9 \frac{V}{\delta_{\nu}}\approx 1.62\frac{V}{\delta_{\nu}}$. 
To obtain a similar magnitude of the normal velocity 
gradient at the plate for the two cases, we should consider an Ekman BL that is approximately 41\% thicker 
than the Blasius BL thickness, thus we should have an Ekman BL thickness $2.284\delta_{Ek}\approx 1.41\delta_{\nu}$. 
Substituting this approximate equality into Eq. (\ref{eq:rossbyBL}), 
yields an enhanced value of $Ro$ by a factor $1.41^2$, thus $Ro_{c_1,th}\approx 2.8$, which is very close to $Ro_{c_1}\approx 2.7$.\\
\indent When $Ro\gtrsim 2.8$ the normal velocity gradient at the wall is determined by the Prandtl-Blasius BL and remains more or less 
constant (and $Nu$ will not change drastically by changing the rotation rate). For sufficiently high 
rotation rates such that $Ro\lesssim 2.8$, we expect that the mean velocity gradient in the BL 
will progressively steepen with increasing rotation rate as the Ekman BL thickness decreases 
proportional to $1/\sqrt{\Omega}$. {\CR{The thermal BL is fully embedded 
within the viscous BL and in case the latter gets thinner the thermal BL is increasingly exposed 
to larger average tangential velocities resulting in an increased normal temperature gradient at 
the plate. We expect that as a consequence also the thickness $\delta_T$ of the thermal BL 
must become somewhat thinner, see, for example, Ref.~\cite{Kunnen2010}. The Nusselt 
number, defined as $Nu=\frac{H}{2\lambda_T}$, should increase}} as heat will be released more efficiently to the (top) plate. Meanwhile, sheet-like thermal plumes 
will be formed at the plates and the rotation rate is not yet sufficiently strong to initiate a strong 
swirl in the converging flow at the base of the sheet-like thermal plume. At the second critical Rossby 
number ($Ro_{c_2}$), the rotation is strong enough to suddenly support a strong swirling flow towards the thermal plumes 
and simultaneously a strong vortical motion is set up in the suddenly emerging vertically-aligned vortex tubes. 
This interpretation is also supported by the joint PDFs of vertical vorticity and horizontal Lagrangian velocity as between both critical Rossby numbers 
a clear change in behavior from LSC-like flow to swirling flows near the top plate for 
decreasing Rossby number is observed. {\COM{Although this scenario has 
to be analyzed in more detail, the main message is that the steepening of the velocity gradients at the wall and 
the generation of swirling converging flows in the BLs are two partially separate processes with different manifestations: 
(i) heat transfer enhancement at $Ro_{c_1}$ and (ii) sudden generation of vertically-aligned vortex tubes connecting 
BL to the bulk at $Ro_{c_2}$ and remarkably not at $Ro_{c_1}$, in contrast to the common 
assumption in the literature.}}\\
\indent In future research, it is of interest to study the link between both 
transitions in more detail by looking at, for example, the dissipation rate at different vertical positions 
around the transition. Since the strength of the thermal plumes is known to depend on the Prandtl number, 
studying the effect of rotation on Lagrangian acceleration statistics at different Prandtl numbers can 
give further insight into the plume dynamics in rotating RBC. 
\begin{acknowledgements}
\small \noindent This work is financially supported by the Nederlandse Organisatie voor Wetenschappelijk 
Onderzoek I (NWO-I), the Netherlands. The authors gratefully acknowledge the support of NWO for the use of 
supercomputer facilities (Cartesius) under Grant No. 16289. R.P.J.K. has received funding from the H2020 
European Research Council (ERC) under the European Union's Horizon 2020 research and innovation 
programme (grant number 678634). EU-COST action MP1305 "Flowing matter" is gratefully acknowledged.
\end{acknowledgements}

\end{document}